\documentclass[twoside,twocolumn,9pt]{article}
\usepackage{extsizes}
\usepackage[super,sort&compress,comma]{natbib} 
\usepackage[version=3]{mhchem}
\usepackage[left=1.5cm, right=1.5cm, top=1.785cm, bottom=2.0cm]{geometry}
\usepackage{balance}
\usepackage{times,mathptmx}
\usepackage{sectsty}
\usepackage{graphicx} 
\usepackage{lastpage}
\usepackage[format=plain,justification=justified,singlelinecheck=false,font={stretch=1.125,small,sf},labelfont=bf,labelsep=space]{caption}
\usepackage{float}
\usepackage{fancyhdr}
\usepackage{fnpos}
\usepackage[english]{babel}
\usepackage{array}
\usepackage{droidsans}
\usepackage{charter}
\usepackage[T1]{fontenc}
\usepackage[usenames,dvipsnames]{xcolor}
\usepackage{setspace}
\usepackage[compact]{titlesec}

\definecolor{cream}{RGB}{222,217,201}

\usepackage{amsmath}
\usepackage{amssymb}
\usepackage{dblfloatfix}

\newcommand{\prl}{Phys.~Rev.~Lett.}
\newcommand{\jcp}{J.~Chem.~Phys.}
\newcommand{\pre}{Phys.~Rev.~E}
\newcommand{\prb}{Phys.~Rev.~B}
\newcommand{\apj}{Astrophys. J.}
\newcommand{\pnas}{Proc.~Nat.~Acad.~Sci.~USA}

\newcommand{\be}{{\bf e}}

\newcommand{\bv}{{\bf v}}
\newcommand{\bu}{{\bf u}}

\newcommand{\br}{{\bf r}}

\newcommand{\qmar}{{\mathcal{G}}}
\newcommand{\qadim}{{g}}

\newcommand{\rcite}[1]{ref.~\citep{#1}}
\newcommand{\rcites}[1]{ref.~\citep{#1}}
\newcommand{\eq}[1]{Eqn~(\ref{#1})}
\newcommand{\eqs}[1]{equations~(\ref{#1})}
\DeclareMathOperator{\sign}{sign}
\newcommand{\fig}[1]{Fig.~\ref{#1}}

\begin{document}

\pagestyle{fancy}
\thispagestyle{plain}
\fancypagestyle{plain}{

  \renewcommand{\headrulewidth}{0pt}
}

\makeFNbottom
\makeatletter
\renewcommand\LARGE{\@setfontsize\LARGE{15pt}{17}}
\renewcommand\Large{\@setfontsize\Large{12pt}{14}}
\renewcommand\large{\@setfontsize\large{10pt}{12}}
\renewcommand\footnotesize{\@setfontsize\footnotesize{7pt}{10}}
\makeatother

\renewcommand{\thefootnote}{\fnsymbol{footnote}}
\renewcommand\footnoterule{\vspace*{1pt}%
\color{cream}\hrule width 3.5in height 0.4pt \color{black}\vspace*{5pt}} 
\setcounter{secnumdepth}{5}

\makeatletter 
\renewcommand\@biblabel[1]{#1}            
\renewcommand\@makefntext[1]%
{\noindent\makebox[0pt][r]{\@thefnmark\,}#1}
\makeatother 
\renewcommand{\figurename}{\small{Fig.}~}
\sectionfont{\sffamily\Large}
\subsectionfont{\normalsize}
\subsubsectionfont{\bf}
\setstretch{1.125} 
\setlength{\skip\footins}{0.8cm}
\setlength{\footnotesep}{0.25cm}
\setlength{\jot}{10pt}
\titlespacing*{\section}{0pt}{4pt}{4pt}
\titlespacing*{\subsection}{0pt}{15pt}{1pt}

\fancyfoot{}
\fancyfoot[RO]{\footnotesize{\sffamily{1--\pageref{LastPage} ~\textbar  \hspace{2pt}\thepage}}}
\fancyfoot[LE]{\footnotesize{\sffamily{\thepage~\textbar\hspace{3.45cm} 1--\pageref{LastPage}}}}
\fancyhead{}
\renewcommand{\headrulewidth}{0pt} 
\renewcommand{\footrulewidth}{0pt}
\setlength{\arrayrulewidth}{1pt}
\setlength{\columnsep}{6.5mm}
\setlength\bibsep{1pt}

\makeatletter 
\newlength{\figrulesep} 
\setlength{\figrulesep}{0.5\textfloatsep} 

\newcommand{\topfigrule}{\vspace*{-1pt}%
\noindent{\color{cream}\rule[-\figrulesep]{\columnwidth}{1.5pt}} }

\newcommand{\botfigrule}{\vspace*{-2pt}%
\noindent{\color{cream}\rule[\figrulesep]{\columnwidth}{1.5pt}} }

\newcommand{\dblfigrule}{\vspace*{-1pt}%
\noindent{\color{cream}\rule[-\figrulesep]{\textwidth}{1.5pt}} }

\makeatother

\twocolumn[
  \begin{@twocolumnfalse}
\vspace{3cm}
\sffamily
\begin{tabular}{m{4.5cm} p{13.5cm} }

& 
  \noindent\LARGE{\textbf{Phase coexistence in a monolayer of active particles induced by Marangoni flows}} 
  \\
  \vspace{0.3cm} & \vspace{0.3cm} \\

 & \noindent\large{Alvaro Dom\'\i nguez,\textit{$^{a}$} Mihail N.~Popescu\textit{$^{b,c}$}} \\

& 
\noindent\normalsize{Thermally or chemically active colloids generate thermodynamic
  gradients in the solution in which they are immersed and thereby
  induce hydrodynamic flows that affect their dynamical evolution.
  Here we study a mean--field model for the many--body dynamics of a
  monolayer of active particles located at a fluid--fluid
  interface. In this case, the activity of the particles creates
  long--ranged Marangoni flows due to the response of the interface,
  which compete with the direct interaction between the particles.
  For the most interesting case of a $r^{-3}$ soft repulsion that
  models the electrostatic or magnetic interparticle forces, we show
  that an ``onion-like'' density distribution
  will develop within the monolayer. For a sufficiently large
  average density, two--dimensional phase transitions (freezing from liquid to
  hexatic, and melting from solid to hexatic) should be observable in a
  radially stratified structure. Furthermore, the analysis allows us to conclude that,
  while the activity may be too weak to allow direct detection
  of such induced Marangoni flows, 
  it is relevant as a collective effect in the emergence of the
  experimentally observable spatial structure of phase
  coexistences noted above. Finally, the relevance of these results
  for potential experimental realizations is critically discussed.} 
  \\


\end{tabular}

 \end{@twocolumnfalse} \vspace{0.6cm}

  ]

\renewcommand*\rmdefault{bch}\normalfont\upshape
\rmfamily
\section*{}
\vspace{-1cm}


\footnotetext{\textit{$^{a}$~F\'\i sica Te\'orica, Universidad de Sevilla, Apdo.~1065, 
41080 Sevilla, Spain. e-mail: \texttt{dominguez@us.es}}}
\footnotetext{\textit{$^{b}$~Max-Planck-Institut f\"ur Intelligente Systeme, Heisenbergstr.~3, 70569 Stuttgart, Germany.}}
\footnotetext{\textit{$^{c}$~IV.~Institut f\"ur Theoretische Physik, Universit\"{a}t Stuttgart, Pfaffenwaldring 57, D-70569 Stuttgart, Germany.}}





\section{Introduction}
\label{sec:intro}

During the last decade the issue of endowing with motility micro- and nano-sized 
particles which are suspended in a liquid  has received significant interest from both 
perspectives of applied and basic science (see the recent reviews in 
\rcites{LaugaRev,ebbens,Gompper2015_rev,Bechinger_RMP2016}). A particularly promising 
approach is the case of chemically or thermally active particles: they can generate 
inhomogeneities in the chemical composition or the temperature of the surrounding 
suspension and achieve self-phoresis by coupling to the gradients of 
these self-generated inhomogeneities 
\cite{Paxton2004,Golestanian2005,Howse2007,Kapral2007}. The motion of such kind of particles, 
either in unbounded fluid, or the vicinity of interfaces, or at interfaces, has been subject 
of numerous theoretical, e.g., \rcites{Golestanian2005,Kapral2007,
Julicher,Popescu2009,Popescu2010,Lowen2011,Seifert2012a,Koplik2013,Brown2017,
Yariv2016,Chang2016,Uspal2015,Ibrahim2016,
Sharifi-Mood2016,Malgaretti2016,Malgaretti2018,Oshanin2017}
and experimental, e.g., 
\rcites{Paxton2004,SenRev,ebbens,Howse2007,Stocco,Howse2015,
Simmchen2016,Uspal2016,Wurger2017,Volpe2018,Bechinger2014,Bechinger2016,Uspal2018a,
Uspal2018b,Uspal2018c} studies. 

The less explored question of the behavior of active particles near, or trapped at, 
liquid-fluid interfaces has been recently tackled both experimentally  
\cite{Stocco,Isa2017,Wurger2017,Isa2018} and theoretically 
\cite{LGN97,Masoud2014,Wurger2014,Stark2014,Malgaretti2016,DMPD16a,DMPD16b,
Malgaretti2018}. An interesting aspect specific to a liquid--fluid interface is that 
the interface itself can  
respond to the chemical or thermal activity of the particles due 
to the locally induced changes in surface tension. These give rise to Marangoni stresses 
which drive hydrodynamic flows in the bulk phases and thus couple back, and influence, 
the motion of the active particles. For example, for a uniformly active spherical particle 
it was shown that the self-induced Marangoni flows can move the particle towards or 
away from the planar interface \cite{LGN97,DMPD16a}. A similar mechanism may set a Janus 
sphere trapped at the interface in motion along the interface 
\cite{Masoud2014,Wurger2014}. Instabilities of interfaces covered by 
active particle, which can be considered themselves to act as surfactants, have been 
reported in \rcite{Stark2014}, while the emergence of collective motion in monolayers of 
active 
particles has been discussed in \rcites{Shelley2014,DMPD16b}. In view of the 
recent experiments described in \rcites{Isa2017,Isa2018,Wurger2017}, it seems that the 
set-up of monolayers of active colloids near or at fluid interfaces such
as, e.g., water--hexadecane, has become feasible and thus various theoretical predictions 
could be eventually tested. 

As discussed in \rcites{DMPD16a, DMPD16b}, the in-plane components of
the Marangoni flows induced by a single particle give rise to
long--ranged effective interactions between pairs located at the
interfacial plane. These interactions provoke collective effects
  even for spherically symmetric particles, that would otherwise ---
  i.e., if isolated --- exhibit no self--propulsion. They can have an
attractive or repulsive character, depending on how the surface
tension reacts to the (chemical or thermal) inhomogeneities. This
effective interaction can compete with the direct forces (like
electrostatic double layer interactions) that the particles exert on
each other.  For example, in \rcite{DMPD16b} the stability of a
monolayer under the competition between the self--induced Marangoni
flows and the capillary attraction was addressed.  In this work, we
consider mutually repulsive particles, that are modeled either as
hard--spheres (e.g., sterically stabilized colloids) or as
``soft--spheres'' exhibiting a long--ranged repulsion (e.g.,
paramagnetic colloidal particles in an external magnetic field as
employed in, e.g., \rcite{ZLM99}, ionizable particles at a
water--dielectric fluid interface as considered in, e.g.,
\rcite{KGHC17}, or polarizable particles in an external electric
field, as described in, e.g., \rcite{ASJN08}). The question we address
is that of the steady-state structure of monolayers located close to,
or at, a liquid fluid interface, formed by such particles when endowed
with thermal or chemical activity.

The paper is organized as follows. In Sec.~\ref{sec:theory} we formulate a conceptually 
simple model for the dynamics of such a monolayer of active particles.
In Sec.~\ref{sec:discussion} we discuss the predictions of this model for 
the two cases of repulsive interactions (hard-- or soft--spheres) noted above, with an 
emphasis on possible experimental realizations. Finally, in Sec.~\ref{sec:conclusions} we 
present our conclusions.

\section{Theoretical model}
\label{sec:theory}

We consider a collection of spherically symmetric, colloidal
particles that form a monolayer because they are constrained to lie at
the flat interface between two fluids. We adopt a coarse--grained
approach, in which the description is based on continuum fields
defined at the monolayer plane. The latter is identified with the
plane $z=0$, so that $\br=(x,y)$ will denote the in-plane position and
\begin{equation}
  \nabla := \left( \frac{\partial}{\partial x},
    \frac{\partial}{\partial y} \right) 
\end{equation}
will denote the two--dimensional (2D) nabla operator in the monolayer plane. The areal
number density of particles in the monolayer is given by the field
$\varrho(\br,t)$. 
Assuming that there is no particle flux in or
out of the monolayer, this field satisfies the continuity equation,
\begin{subequations}
  \label{eq:model}
\begin{equation}
  \label{eq:cont}
  \frac{\partial \varrho}{\partial t} = - \nabla\cdot (\varrho \bv) ,
\end{equation}
where $\bv(\br,t)$ is the monolayer velocity field. This velocity is
driven by the gradient of the chemical potential $\mu(\varrho)$ of the
monolayer (the ``thermodynamic'' force) \cite{Batc76,BDO15} and by
the drag due to the three--dimensional (3D), incompressible ambient
flow in the surrounding fluids. 
In the overdamped approximation
(see, e.g., \rcite{Dhon96}), the velocity field is given by
\begin{equation}
  \label{eq:v}
  \bv = - \Gamma\nabla\mu + \bu ,
\end{equation}
where $\Gamma$ is the mobility of the particles in the monolayer,
and $\bu(\br)$ is the 3D Marangoni flow \emph{evaluated at the
    monolayer plane.} This flow 
is, in turn, induced by the chemical activity
of the particles: each particle alters the chemical composition or the temperature field 
of the environment, which induces local changes in the properties of the
fluid interface. In particular, the gradients in surface tension
(Marangoni stresses) set the ambient fluids in motion: the
corresponding Marangoni flow in an unbounded fluid and evaluated at the plane $z=0$ is 
given as \cite{LGN97,MaSh14,Wuer14,DMPD16a}
\begin{equation}
  \label{eq:u}
    \bu(\br) = \frac{\qmar}{2\pi}\int d^2\br' \; \varrho(\br')
    \frac{\br-\br'}{|\br-\br'|^2}
    = - \nabla \Phi,
  \end{equation}
  \begin{equation}
    \Phi(\br) = - \frac{\qmar}{2\pi} \int d^2\br' \; \varrho(\br')
    \ln |\br-\br'| .
  \end{equation}
\end{subequations}
Here the factor $\qmar$, that can be positive or negative, characterizes the 
magnitude of the induced Marangoni flow (see \eq{eq:Qdef} in 
Sec.~\ref{sec:discussion}). One can notice that \eq{eq:u} is formally identical with a 
2D Newtonian gravitational field if $\qmar<0$ (or ``antigravitational'' when
$\qmar>0$); the corresponding ``2D field equation'' is
\begin{equation}
  \label{eq:gravity}
  \nabla^2\Phi = - \nabla\cdot\bu = - \qmar \varrho .
\end{equation}
Note that while the 3D ambient flow is incompressible, the 2D flow $\bu(\br)$, i.e., 
evaluated at the plane of the monolayer is compressible (its 2D divergence is 
non-vanishing). Thus, the Marangoni flow plays the role of a collective attraction (when 
$\qmar<0$) or repulsion (if $\qmar>0$). One could indeed interpret the dynamics of the 
model described by \eqs{eq:model} as a self--gravitating 2D fluid of interacting Brownian 
particles.

Equations~(\ref{eq:model}) build a complete model from which the dynamical
evolution of the particle distribution $\varrho(\br,t)$ in the
monolayer can be obtained. The dynamics is intrinsically
  \emph{collective}; in particular, an isolated particle is not
  self--propelled and would remain at rest due to the spherical
  symmetry. This very simplified model should capture the interplay
between the direct interparticle forces (described by the chemical
potential) and the chemical activity (via the Marangoni flow). The
physical assumptions and simplifications involved in this model have
been discussed elsewhere in detail \cite{DMPD16a, DMPD16b} and
therefore here we only succinctly recall them. In essence,
\eqs{eq:model} provide a coarse--grained description, valid for
sufficiently large scales and long times, for a sufficiently dilute
monolayer. Thus, \eq{eq:v} assumes overdamped motion driven by the
chemical potential (hypothesis of local equilibrium), and by the
drag. The latter is given by the Stoke's equations (low Reynolds and
Mach numbers \cite{Dhon96}) within the point--particle approximation
\cite{DMPD16a}, so that \eq{eq:u} represents actually a
mean--field-like approximation to the long--ranged Marangoni flow in
the dilute limit. One neglects the effect of interparticle
hydrodynamic interactions as compared to the drag by the Marangoni
flow, although they could be accounted for within the coarse--grained
description without significant conceptual difficulties: the
short--separation contributions would appear as a density dependence
of the mobility $\Gamma$ \cite{Nozi87}, while the long--range
contribution, which induces collective anomalous diffusion in the
monolayer \cite{BDO15, BDO16}, would show up as an additional drag
flow in \eq{eq:v}. Regarding the source of the Marangoni flow, i.e.,
inhomogeneities in the concentration of chemicals or in the ambient
temperature field, it is assumed to be always in equilibrium with the
instantaneous particle configuration (fast--relaxation
approximation). Finally, for reasons of simplicity we neglect
complicated rheological properties of the interface; only its surface
tension matters. This also concerns its role as a passive constraint
for the colloidal particles, e.g., when the latter are partially
wetted by the fluids and thus get trapped by strong wetting forces
(see, e.g., \rcite{Bink02}).

\begin{figure*}[b]
  \centering 
  \includegraphics[width=.45\textwidth]{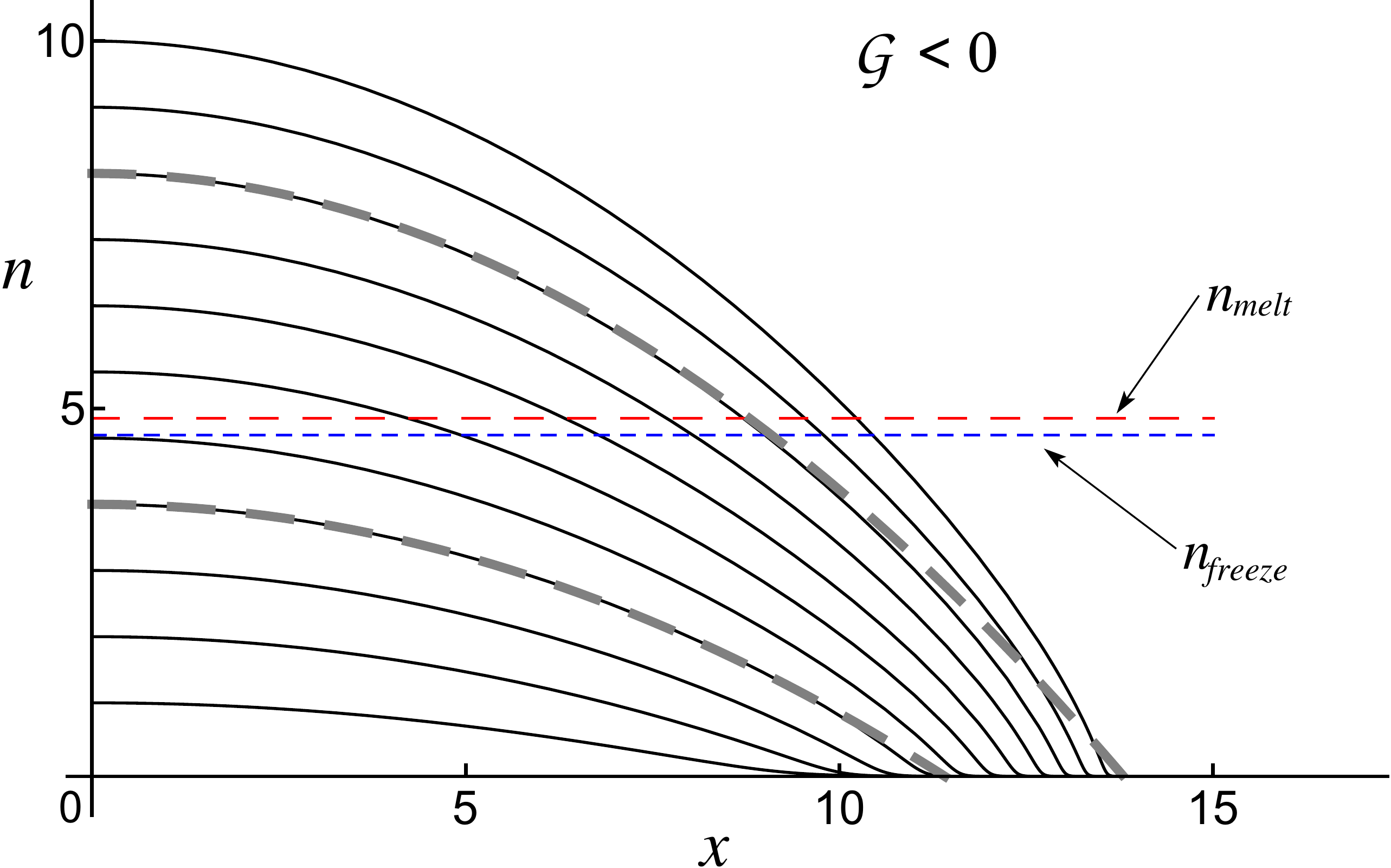}
  \hfill
  \includegraphics[width=.45\textwidth]{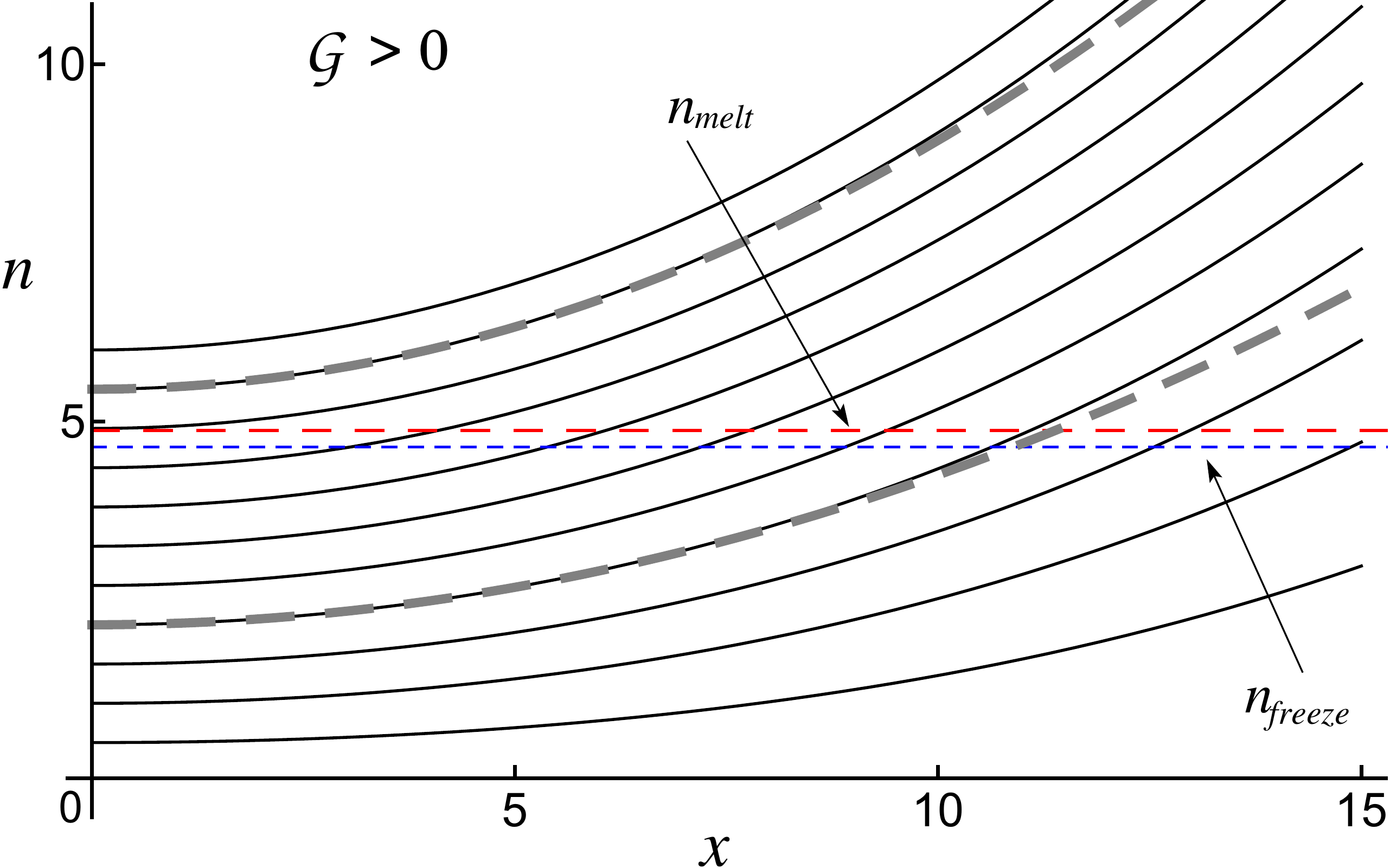}
   \caption{Typical form of the density profile $n(x; n_0)$ (solid lines) of 
   ``soft spheres''
    for ``attractive'' ($\qmar < 0 $, left panel) or ``repulsive'' ($\qmar > 0 $, 
    right panel) Marangoni flow. The horizontal lines correspond to the 
    values $n_\mathrm{freeze}$ and $n_\mathrm{melt}$, at which the phase
    transitions occur. The dashed 
    lines show, for each of the cases 
    $\qmar \gtrless 0$, two examples of the parabolic approximation given by 
    \eq{eq:approxsoftprofile}.}
  \label{fig:softprofile}
\end{figure*}

The stationary state of the monolayer ($\partial\varrho/\partial t=0$) is given by the 
condition
\begin{equation}
  \label{eq:stat1}
  \bv= 0 
  \qquad\Rightarrow\qquad
  - \Gamma \nabla\mu + \bu = 0 .
\end{equation}
Since $\bu$ is given by \eq{eq:u}, this is an integrodifferential equation for the 
stationary profile $\varrho(\br)$. However, the mathematical analogy with gravity, 
expressed by \eq{eq:gravity}, allows one to reduce \eq{eq:stat1} to a partial differential
equation for the profile: taking the divergence of \eq{eq:stat1} renders
\begin{equation}
  \label{eq:stat2}
  - \Gamma \nabla^2\mu(\varrho) + \qmar \varrho = 0 . 
\end{equation}
This equation must be complemented by the appropriate boundary conditions. The form of 
the Marangoni flow~(\ref{eq:u}) assumes that the fluid occupies an unbounded domain, 
i.e., we assume implicitly that there is no relevant influence by the distant 
boundary conditions and look, therefore, for radially symmetric solutions expressing 
spatial isotropy. In experiments, for instance, one can think of a monolayer confined laterally by a distant 
circular hard wall. Alternatively, one can also consider a circular optical trap or 
a coarse circular sieve that confines the particles only without disrupting appreciably 
the ambient flow or the sources of the Marangoni flow (distribution of chemicals or 
temperature gradients). It is thus still meaningful to study solutions to \eq{eq:stat2} 
for a monolayer of a finite spatial extent, say $r< {\cal L}$, while using 
the expression (\ref{eq:u}) for flow in an unbounded region. 

\begin{figure*}[b]
  \centering   
  \includegraphics[width=.45\textwidth]{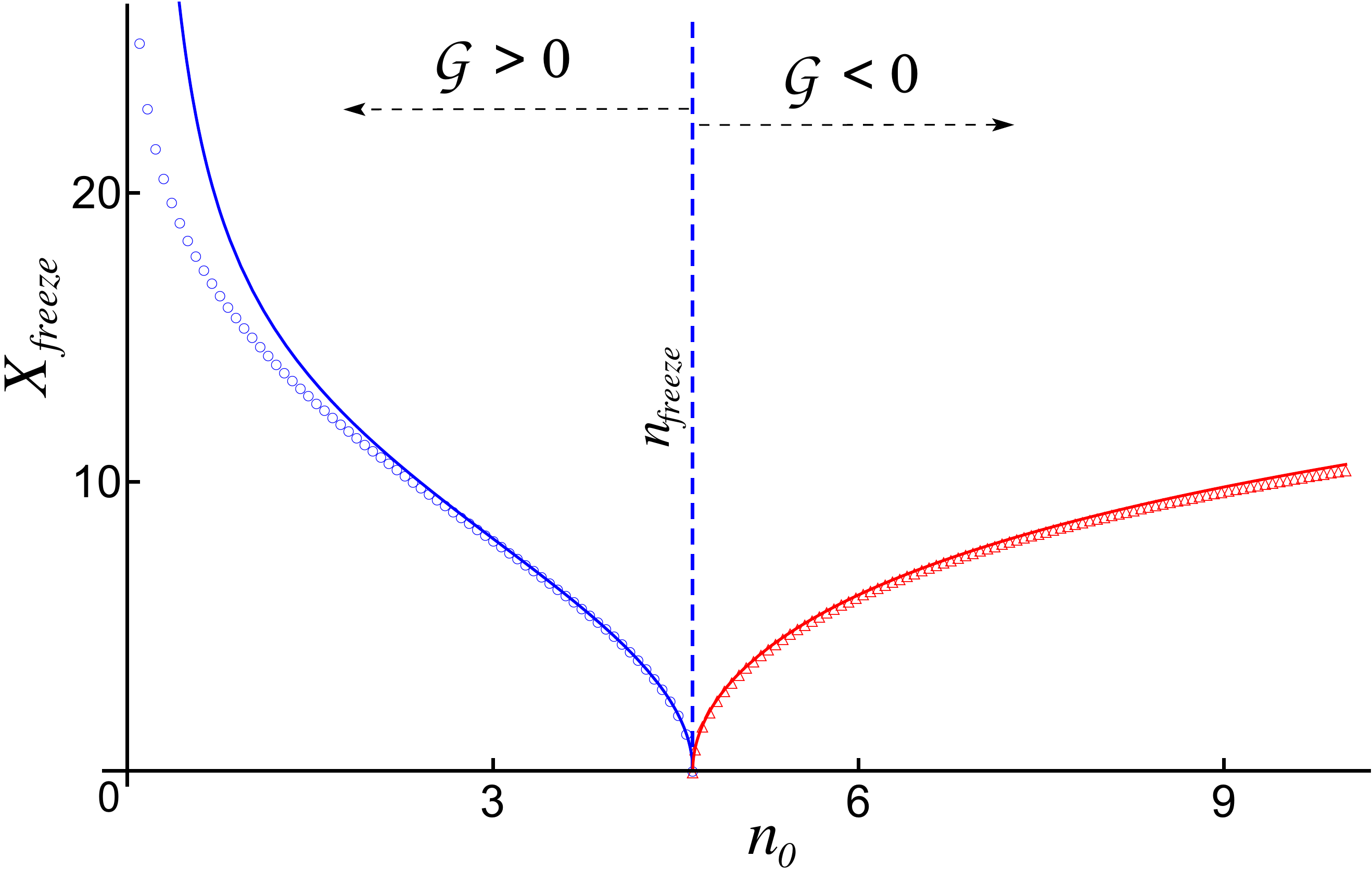}
  \hfill
  \includegraphics[width=.45\textwidth]{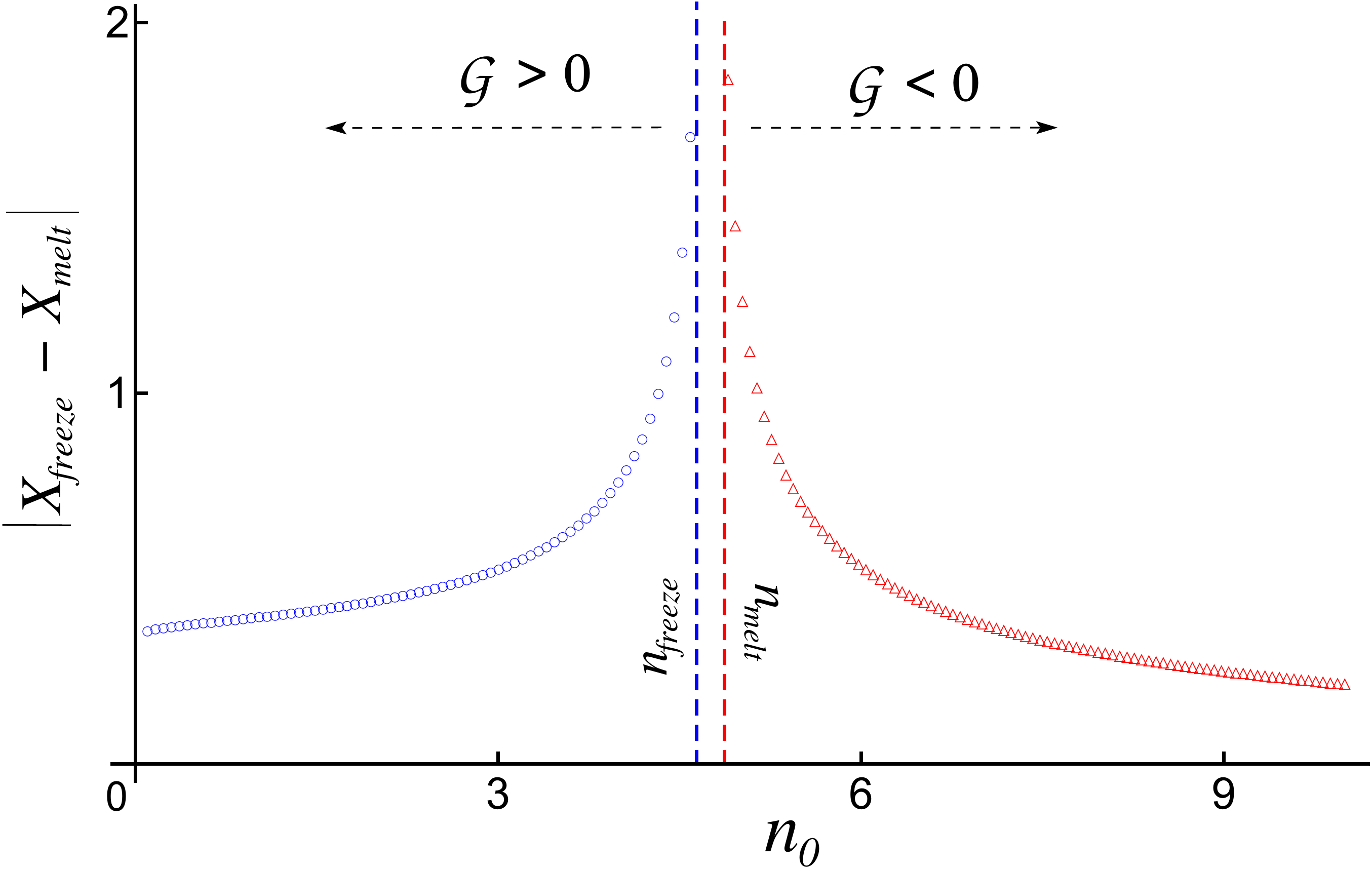}
  \caption{(Left) The value of the radial distance $X_\mathrm{freeze}$
    from the center at which the freezing transition in ``soft spheres''
    occurs for each profile (labelled by the value $n_0$ of the
    central density). The solid line is the approximation given by
    \eq{eq:approxXfreeze}. (Right) Plot of the width
    $|X_\mathrm{freeze}-X_\mathrm{melt}|$ of the region where the
    hexatic phase occurs. In both plots, the branch with
    $n_0>n_\mathrm{freeze}$ corresponds to the case of ``attractive''
    Marangoni flow; the other branch is associated to the
    ``repulsive'' case.}
  \label{fig:Xfreeze}
\end{figure*}

 Irrespective of the exact form of these distant boundary conditions, we look 
for radially symmetric solutions, for which \eq{eq:stat2} takes the form
\begin{equation}
  \label{eq:stat3}
  \left(\frac{d^2}{d r^2}+ \frac{1}{r} \frac{d}{d r} \right) \mu(\varrho(r)) =
  \frac{\qmar}{\Gamma} \varrho(r) .
\end{equation}
We look for solutions that are regular (continuous and differentiable); this imposes 
additional boundary conditions at the origin of the coordinate
system, namely,
\begin{equation}
  \label{eq:bc0}
  \varrho(r=0) = \varrho_0 \;\mathrm{finite} ,
  \qquad
  \frac{d\varrho}{d r}(r=0) = 0 .
\end{equation}
Let $T$ denote the temperature ($k$ will be Boltzmann's constant) and
$\varrho_c$ denote a characteristic density of the phases described by
the chemical potential $\mu$. We introduce the dimensionless magnitude
\begin{equation}
  \label{eq:smallq}
  \qadim := \frac{\qmar}{\Gamma k T} ,
\end{equation}
which pertains only to the hydrodynamical effects, and 
the length scale
  \begin{equation}
    \label{eq:lengthscale}
    \ell := \frac{1}{\sqrt{|\qadim| \varrho_c}} ,
  \end{equation}
  which represents the characteristic mean interparticle separation
  $1/\sqrt{\varrho_c}$ rescaled by the hydrodynamic
  factor~(\ref{eq:smallq}). With the definitions
\begin{equation}
  \label{eq:adim}
  x := \frac{r}{\ell} ,
  \qquad 
  n := \frac{\varrho}{\varrho_c},
  \qquad
  \hat{\mu}(n) := \frac{\mu(\varrho)}{k T} ,
\end{equation}
of dimensionless variables, we arrive finally at an initial value problem for an ordinary
differential equation:
\begin{subequations}
  \label{eq:adimpde}
\begin{equation}
  \label{eq:adimstat}
  \left( \frac{d^2}{d x^2}+ \frac{1}{x} \frac{d}{d x} \right) \hat{\mu}(n(x)) =
  \sign{(\qmar)} n(x) ,
\end{equation}
\begin{equation}
  \label{eq:adimbc0}
  n(x=0) = n_0 \;\mathrm{finite} ,
  \qquad
  \frac{dn}{d x}(x=0) = 0 .
\end{equation}
\end{subequations}
This determines the density profile $n(x; n_0)$, which is actually
  identified uniquely by the value $n_0$ of the central density. From
  this profile, the total amount of particles contained within 
a disk of a given radius ${\cal L}$ is calculated as
\begin{subequations}
  \label{eq:cumulativemass}
  \begin{equation}
    \label{eq:M}
    M({\cal L}) := 2\pi \int_0^{\cal L} dr\; r \varrho(r) 
    = \frac{2\pi}{|\qadim|} N(X) ,
  \end{equation}
  where 
  \begin{equation}
    \label{eq:N}
    N(X; n_0) := \int_0^X dx\; x \, n(x; n_0) ,
    \qquad
    X := \frac{\cal L}{\ell} .
  \end{equation}
\end{subequations}
The Marangoni flow, which has the radial direction $\be_r$, can then be computed by a 
straightforward application of Gauss theorem to \eq{eq:gravity},
\begin{equation}
  \label{eq:radialu}
  \bu(\br) = \frac{\qmar M(r)}{2\pi r} \be_r .
\end{equation}
When this expression is combined with the stationarity condition~(\ref{eq:stat1}) and 
the definition~(\ref{eq:M}), one gets
\begin{equation}
  N(X; n_0) = \sign{(\qmar)} X \; \frac{d\hat{\mu}(n(X; n_0))}{dn}
  \; \frac{d n(X; n_0)}{dX}.
\end{equation}
This expression provides a relationship between the three parameters $n_0$, $N$ and $X$ 
(that is, $\varrho_0$, $M$, and ${\cal L}$ in physical variables). Therefore, 
concerning an 
experimental realization, any stationary state can be characterized by the measurement 
either of the central density $\varrho_0$ or of the total number of particles $M$ within 
a 
finite circular region of radius ${\cal L}$. Alternatively, when preparing the 
system 
in the 
laboratory, one can fix the amount $M$ of particles in a region defined by its size 
${\cal L}$ 
(which may be significantly easier to control than the value of $\varrho_0$). This 
flexibility in choosing the variables is particularly important for connecting the 
theoretical analysis with the experimentally accessible quantities since, from the point 
of view of numerical analysis, the solution $n(x; n_0)$ of \eqs{eq:adimpde} is most 
conveniently parametrized by the value $n_0$ of the density at the center of the 
domain.

\section{Results and discussion}
\label{sec:discussion}

We are now in a position to apply the generic framework developed
above to specific cases that are experimentally relevant. Since
  the Marangoni flow can be formally analogous to Newtonian gravity,
  the stationary profiles that solve \eq{eq:stat2} would be the
  equivalent of a material cluster in equilibrium under its own
  gravity. The thermodynamic properties of the monolayer enter via
the chemical potential $\hat{\mu}(n)$. If this includes the existence
of different equilibrium phases, one can then easily envision the emergence
of a radially stratified structure, much like in the postulated
interior of condensed astrophysical objects (planets and stars).

We therefore will consider the solution to \eqs{eq:adimpde} for different choices 
of the chemical potential. We shall focus on two extreme cases that intend to 
describe experimentally relevant systems. In one extreme, there is the case of ``soft 
spheres'', in which particles experience a mutual long--ranged repulsive interaction. 
In the opposite extreme there is the limit of hard spheres, for which the repulsion 
is very short--ranged.

\begin{figure*}
  \centering
   \includegraphics[width=.45\textwidth]{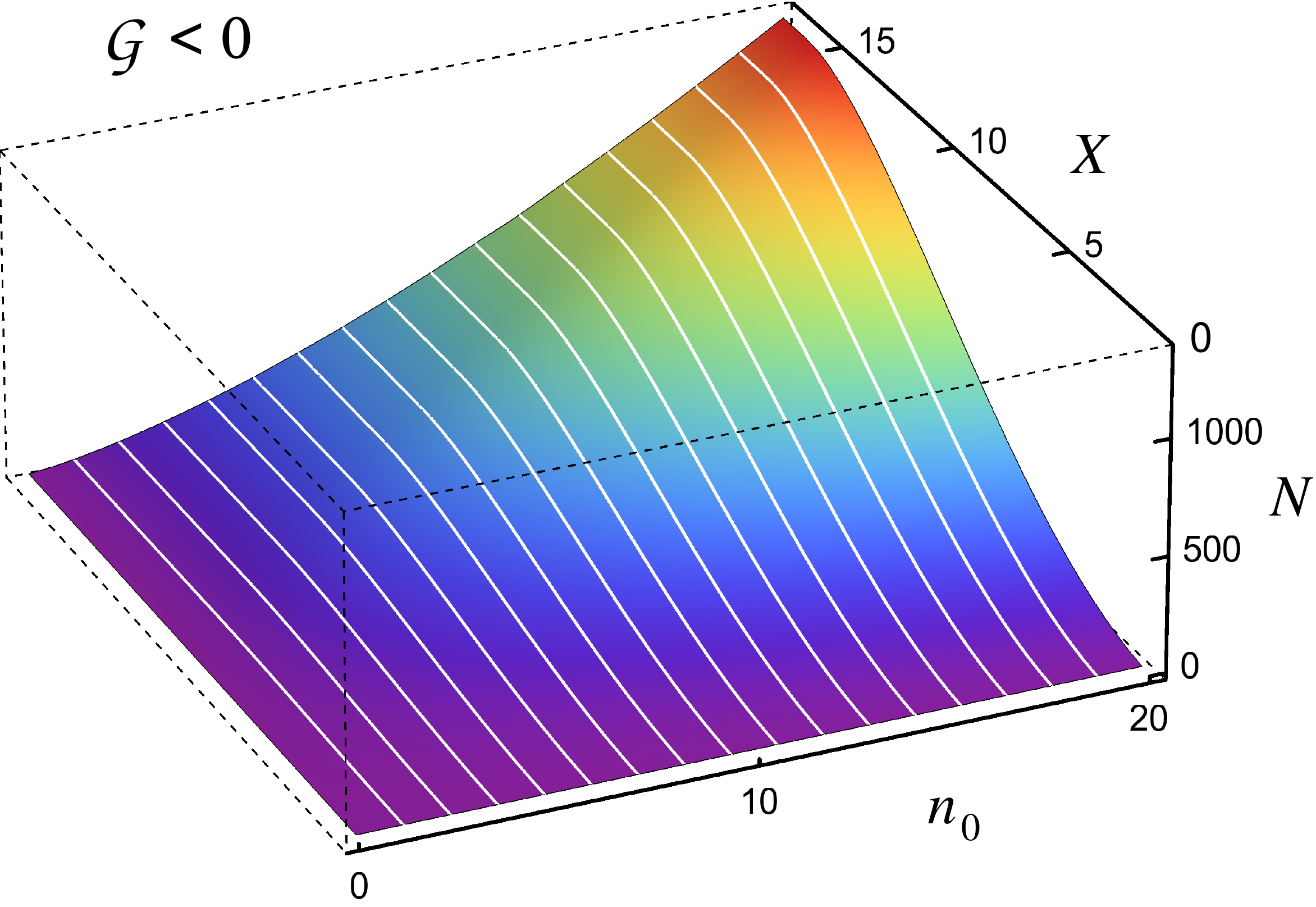}
  \hfill
  \includegraphics[width=.45\textwidth]{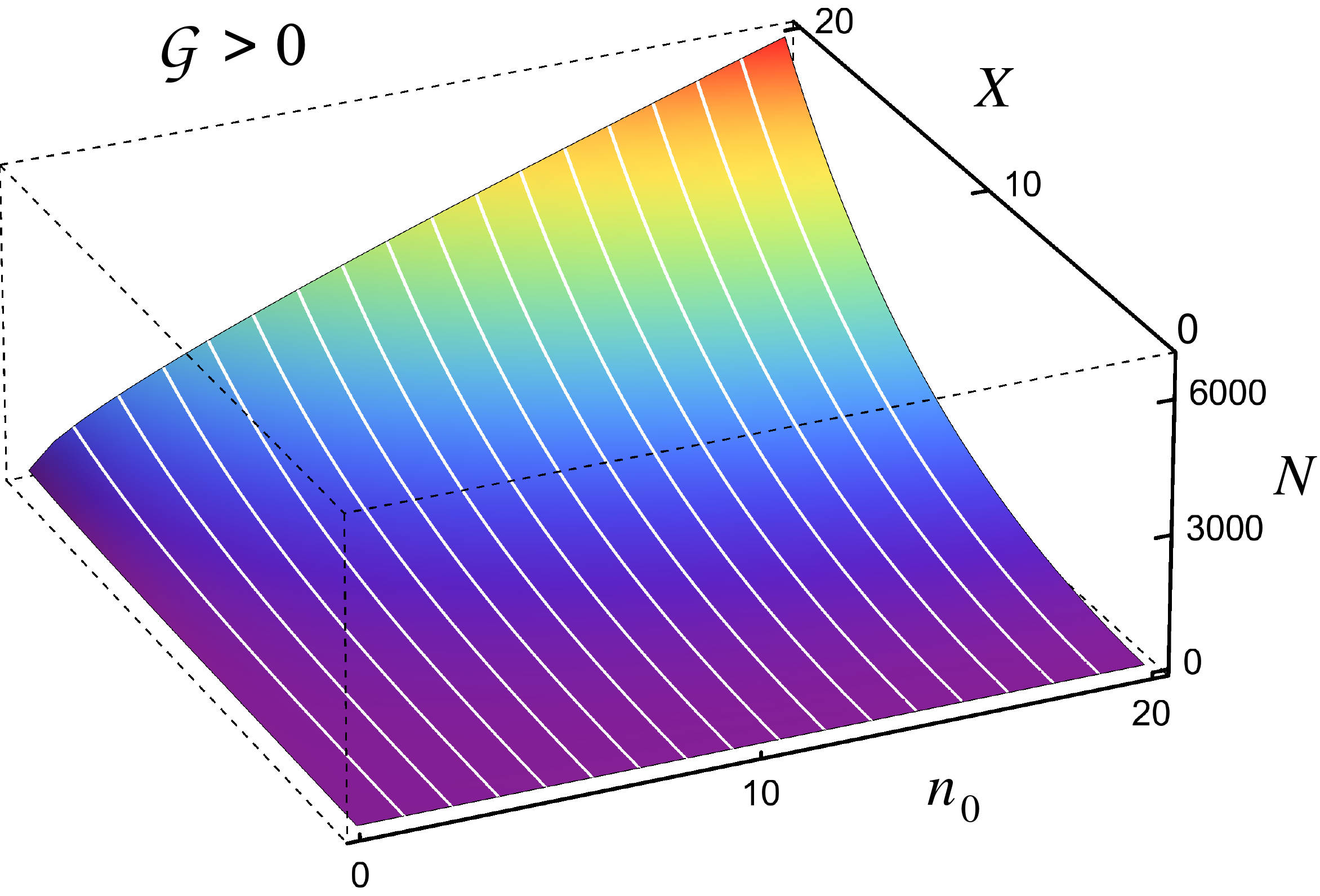} 
  \caption{The function $N(X; n_0)$, i.e., the total amount of
    ``soft spheres'' inside a circular region of given radius, for
    ``attractive''  ($\qmar < 0 $, left panel) or ``repulsive'' 
    ($\qmar > 0 $, right panel) Marangoni
    flow. Each white line corresponds to a curve $N(X)$ for
      a given constant value of $n_0$.
    }
    \label{fig:softN}
\end{figure*}

\subsection{Soft spheres}

 As a first case study, we model the colloidal particles as
``soft--spheres'', a term that characterizes particles that interact
with each other by means of a repulsive power--law potential. More
precisely, we consider a pair potential of the form
\begin{equation}
  \label{eq:Vdip}
  V_\mathrm{dip}(r) = \frac{B}{r^3} = k T \left(\frac{\zeta}{r}\right)^3 ,
\end{equation}
where the constant $B$ --- or equivalently, the parameter
  $\zeta = (B/k T)^{1/3}$ (Bjerrum length) --- characterizes the strength
of the interaction. This potential can model the dipolar repulsion due
to the induced moments in paramagnetic particles immersed in an
external magnetic field \cite{ZLM99}.
It can also describe the asymptotic repulsion
between electrically charged particles when one of the fluids is a
dielectric \cite{Stil61,Hurd85,ACNP00,FDO07,DFO08,PLMB15}, or between polarized
  particles in an external electric field \cite{ASJN08}. The
evaluation of the partition function leads straightforwardly to the
simple scaling form $\mu(\varrho,T) = kT \hat{\mu}(\varrho \zeta^{2})$
of the chemical potential; this provides the function $\hat{\mu}(n)$
in \eq{eq:adim} with the natural choice $\varrho_c := \zeta^{-2}$ as
characteristic density (see Appendix~\ref{app:musoft}). This scaling form implies
that the phase diagram is particularly simple because it does not
depend on two independent control parameters (density and
temperature). Instead, the only relevant parameter is the combination
given by the rescaled density $n=\varrho \zeta^{2}$. 

Particles interacting with \eq{eq:Vdip} are known to exhibit a
discontinuous liquid--solid phase transition in 3D \cite{HRJH70}. In
2D, however, it was predicted (the
Kosterlitz--Thouless--Halperin--Nelson--Young theory
\cite{KoTh73,Youn79,NeHa79}) that a hexatic phase exists, so that
freezing occurs continuously in the variable $n$ in two stages, first
a liquid--hexatic, and then a hexatic--solid transition. This scenario
has been confirmed recently in experiments involving monolayers of
paramagnetic particles \cite{ZLM99} and monolayers of electrically
charged particles \cite{KGHC17}. The range of existence of the hexatic
phase is very narrow: according to the experimental values quoted by
\rcite{ZLM99}, the liquid--hexatic transition (``liquid freezes'')
occurs at $n_\mathrm{freeze} \approx 4.65$, while the hexatic--solid
transition (``solid melts'') takes place at
$n_\mathrm{melt} \approx 4.87$ (the corresponding values quoted in
\rcite{KGHC17} are somewhat larger; this may be due to uncertainties
in the determination of the value of $\zeta$ in \eq{eq:Vdip}). 

The density profile $n(x; n_0)$ is obtained by solving \eqs{eq:adimpde} 
numerically with the fit to the chemical potential $\hat{\mu}(n)$ given by 
\eq{eq:fitmusoft}. \fig{fig:softprofile} shows a representative set of profiles. For not 
too small values of the central density $n_0$, we find that, in the region of 
nonvanishing densities, the profile is very well approximated by its parabolic 
approximation about the center:
\begin{equation}
  \label{eq:approxsoftprofile}
  n(x; n_0) \approx n_0 \left[ 1 + \sign{(\qmar)}
    \left(\frac{x}{\lambda}\right)^2 \right] ,
  \qquad
  \lambda(n_0) := 2\sqrt{\frac{d\hat{\mu}}{dn}(n_0)} .
\end{equation}
(The coefficient $\lambda$ is straightforwardly obtained by looking for a 
solution to the differential equation~(\ref{eq:adimstat}) in the form of a 
Taylor expansion.)
When $\qmar<0$ (``attractive'' Marangoni flow), the density increases
towards the center of the distribution (as for an astrophysical object
under its own gravity). 
One thus expects to observe different local structures, corresponding
to the phase associated with the local value of $n$,
in an ``onion-like''
assembly. So, there would be crystalline order near the center,
surrounded by a disordered isotropic fluid; and between both phases,
there should appear a hexatic shell exhibiting orientational, but not
positional, order. The trend is reversed when $\qmar>0$ (``repulsive''
Marangoni flow).

Another interesting feature is that in the ``atractive'' case the
system is ``self--confined'' in the following sense. Since the density
decays steadily to zero, the farther the external walls are located,
the smaller (vanishing) is the confining pressure exerted by them.
This is a direct consequence of the long--ranged nature of the
Marangoni flow given by \eq{eq:u}: in the language of Statistical
Mechanics, this (``gravitational'') force is nonintegrable and,
provided the system is large enough, dominates over the short--ranged
dipolar repulsion given by \eq{eq:Vdip}. In the opposite case of
``repulsive'' Marangoni, it is obvious that the walls are required to
confine the system against the Marangoni and dipolar repulsions: this
explains the formation of a fluid (low density) central region in the
profiles shown in \fig{fig:softprofile} and a crystalline (high
density) structure at the outer border.

In \fig{fig:Xfreeze} we show the radial distance $X_\mathrm{freeze}(n_0)$ from the 
center at which the freezing transition is expected to occur as a function of the central 
density $n_0$, i.e., the solution of $n(X_\mathrm{freeze}; n_0)=n_\mathrm{freeze}$. The 
prediction given by the
approximated profile~(\ref{eq:approxsoftprofile}), namely
\begin{equation}
  \label{eq:approxXfreeze}
  X_\mathrm{freeze}(n_0) \approx \lambda(n_0)
  \left|\frac{n_0}{n_\mathrm{freeze}}-1\right|^{1/2} ,
\end{equation}
provides an excellent fit for any central density $n_0>2$, and
captures the divergence as $n_0\to 0$. Similarly, one defines the
radial distance $X_\mathrm{melt}(n_0)$ at which the melting transition
occurs. Also shown in \fig{fig:Xfreeze} is the width
$|X_\mathrm{freeze}-X_\mathrm{melt}|$ of the shell where the hexatic
phase would be observed.

Starting from the profiles $n(x; n_0)$ parametrized by the value of
$n_0$, one can compute the function $N(X;n_0)$ given by
\eq{eq:N}. Figure~\ref{fig:softN} shows this function for different
values of $n_0$. In the range of parameters explored, the curves do
not cross and seem to cover the whole $(X,N)$ plane. This means that
the preparation of a system with given $N$ and $X$ determines $n_0$
uniquely, and that all the combinations of the parameters $N$ and $X$
lead to a solution of \eqs{eq:adimpde}, i.e., to a stationary
state. This is not a trivial statement: for instance, it is long
known in the astrophysical community that a self--gravitating ideal
gas (the so-called ``isothermal sphere'' configuration) can lack
stationary states, depending on the values of the parameters (see,
  e.g., the review work \cite{Padm90}). However, this behavior is
usually seen as a limitation of the ideal gas approximation. Our
numerical results for particles with the interaction
potential~(\ref{eq:Vdip}) confirm this expectation.

\begin{figure*}
  \centering 
   \includegraphics[width=.45\textwidth]{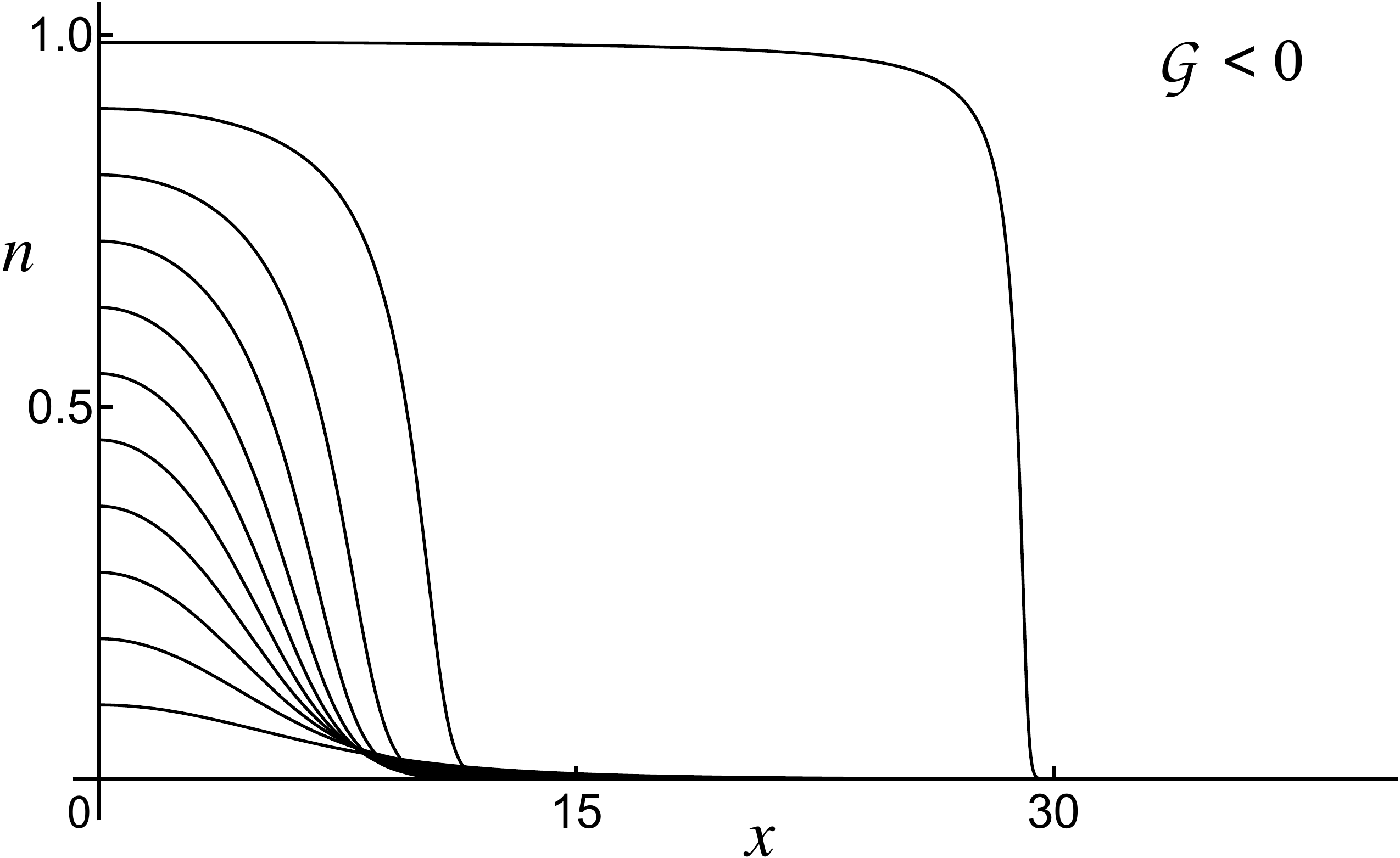}
  \hfill
  \includegraphics[width=.45\textwidth]{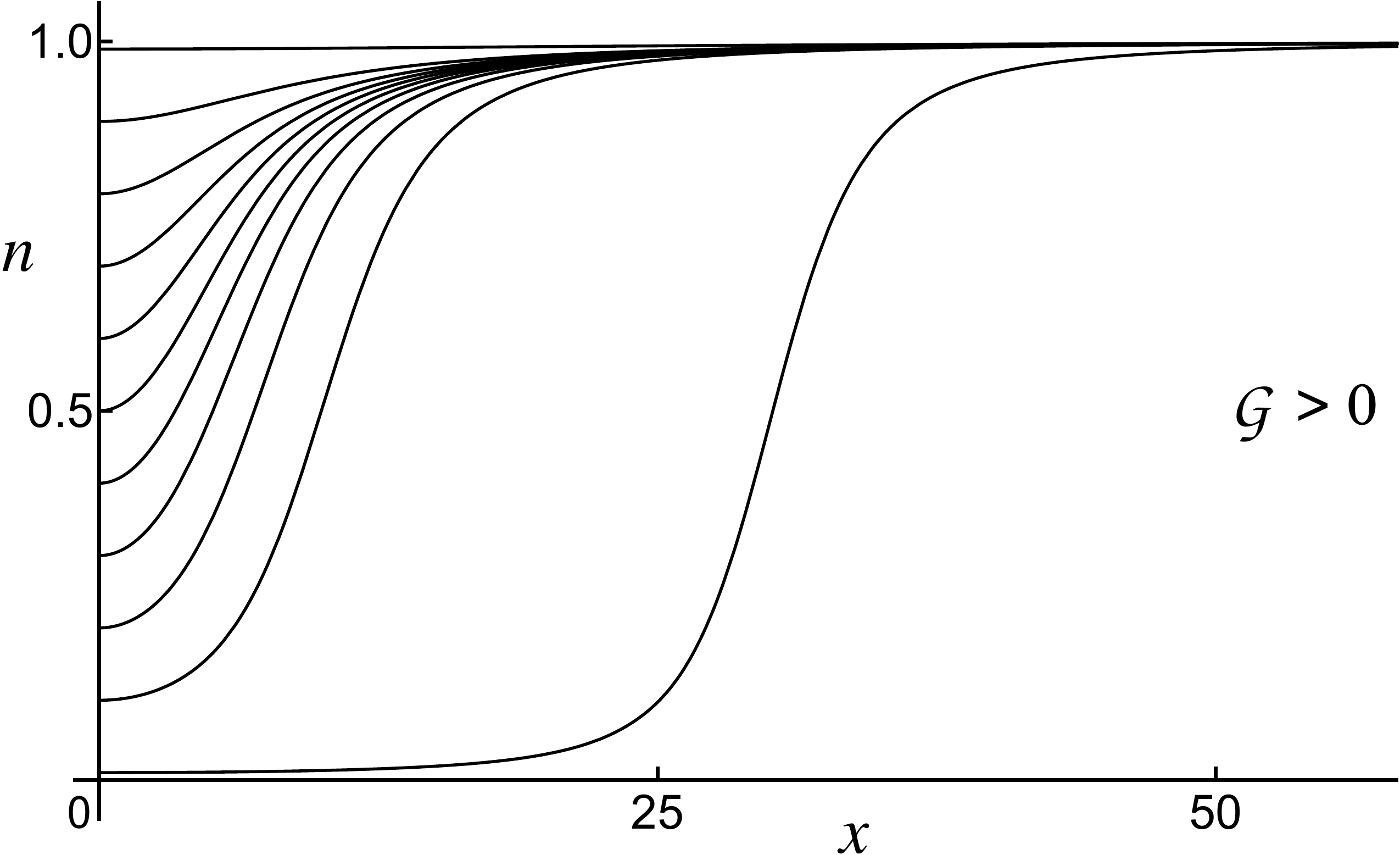} 
  \caption{Typical density profiles $n(x; n_0)$ of hard spheres for
    ``attractive'' (left) or ``repulsive'' (right) Marangoni flow.}
  \label{fig:hardprofile}
\end{figure*}

\subsection{Hard spheres}
\label{sec:hard}

We now consider the case that the colloidal particles are approximated
as ``hard--spheres'', i.e., their only interaction is hard--core
exclusion. The chemical potential for this system also exhibits a
simple scaling (see Appendix~\ref{app:muhard}),
$\mu(\varrho, T) = k T \hat{\mu}(\varrho/\varrho_c)$, with the
close--packing number density $\varrho_c=1/(2\sqrt{3} R^2)$ of
disks of radius $R$. This system does not exhibit any phase
transition, staying in a fluid phase for any density
$n=\varrho/\varrho_c<1$. The limiting value $n=1$ can only be reached
under infinite pressure and corresponds to the formation of a crystal
of hard disks in contact.

\begin{figure}
  \centering    
  \includegraphics[width=.45\textwidth]{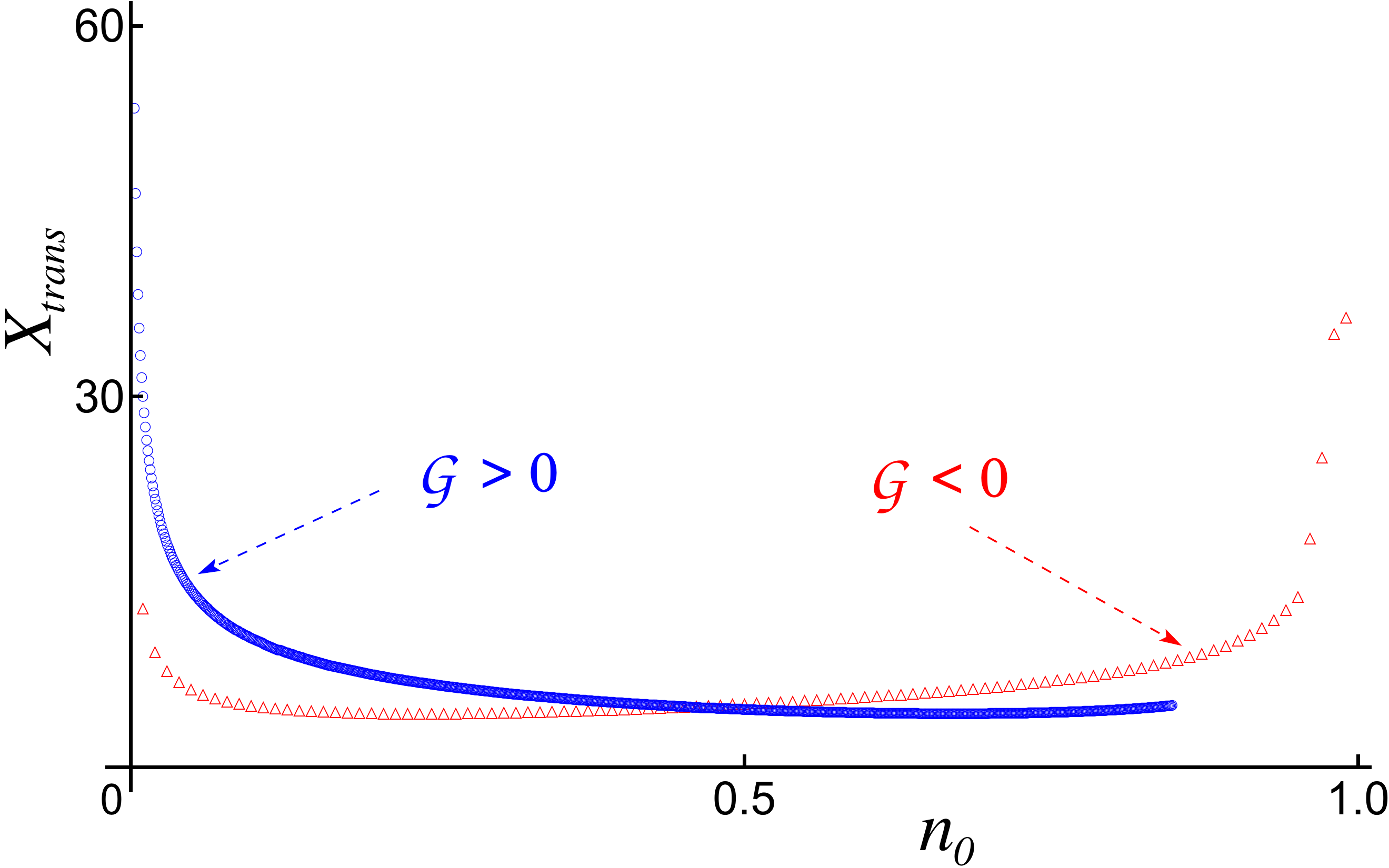}
  \caption{The position of the pseudotransition in the case of hard
    spheres as a function of the central density.}
  \label{fig:Xtrans}
\end{figure}

The density profiles $n(x; n_0)$, shown in \fig{fig:hardprofile}, are
obtained by solving \eqs{eq:adimpde} numerically with the fit to the chemical
potential $\hat{\mu}(n)$ given by \eq{eq:fitmuhard}. The profiles for which
$n(x; n_0)$ is never close to one describe a smoothly varying
distribution. However, for those which come close to $n=1$ (either at
the center for ``attractive'' Marangoni flow or far from it for
``repulsive'' Marangoni flow), the most salient feature is an apparent
phase segregation: a closely--packed crystalline structure emerges in
coexistence with a dilute ``gas phase''. (The exactly close packed
structure with $n=1$ is never reached, but the difference would be
visually unnoticeable in an experimental realization.) The transition
region between both phases has a nonvanishing width (as remarked, the
hard--disk system does not have a true phase transition in the
thermodynamic sense), but the width can be substantially smaller than
the extension of the ``crystal'' and the ``gas''. This would give the
impression of a pseudotransition between two phases. For this kind of
configurations, we have found that the radial position
$X_\mathrm{trans}$ of this pseudotransition can be defined
conventionally by the location of the inflection point of the density
profile, i.e., $n''(X_\mathrm{trans}; n_0)=0$. In
\fig{fig:Xtrans} we show this position as a function of the central
density. We remark on passing that these profiles would correspond
  to the core--halo structures reported in the study of the
  equilibrium configurations of a self--gravitating 3D gas of hard
  spheres \cite{Padm90,ArHa72,SKS95}.
 
\begin{figure*}
  \centering   
  \includegraphics[width=.45\textwidth]{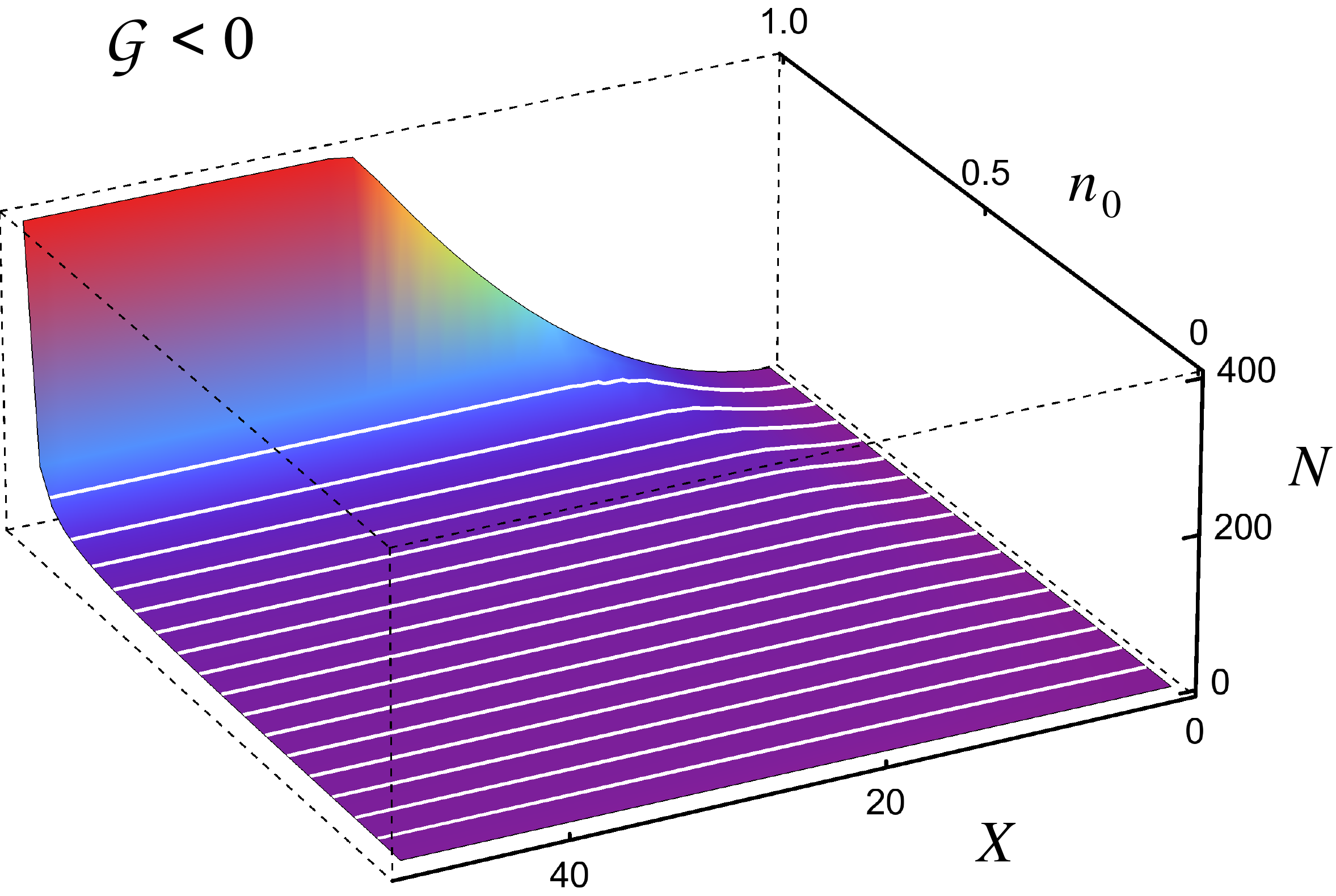}
  \hfill
  \includegraphics[width=.45\textwidth]{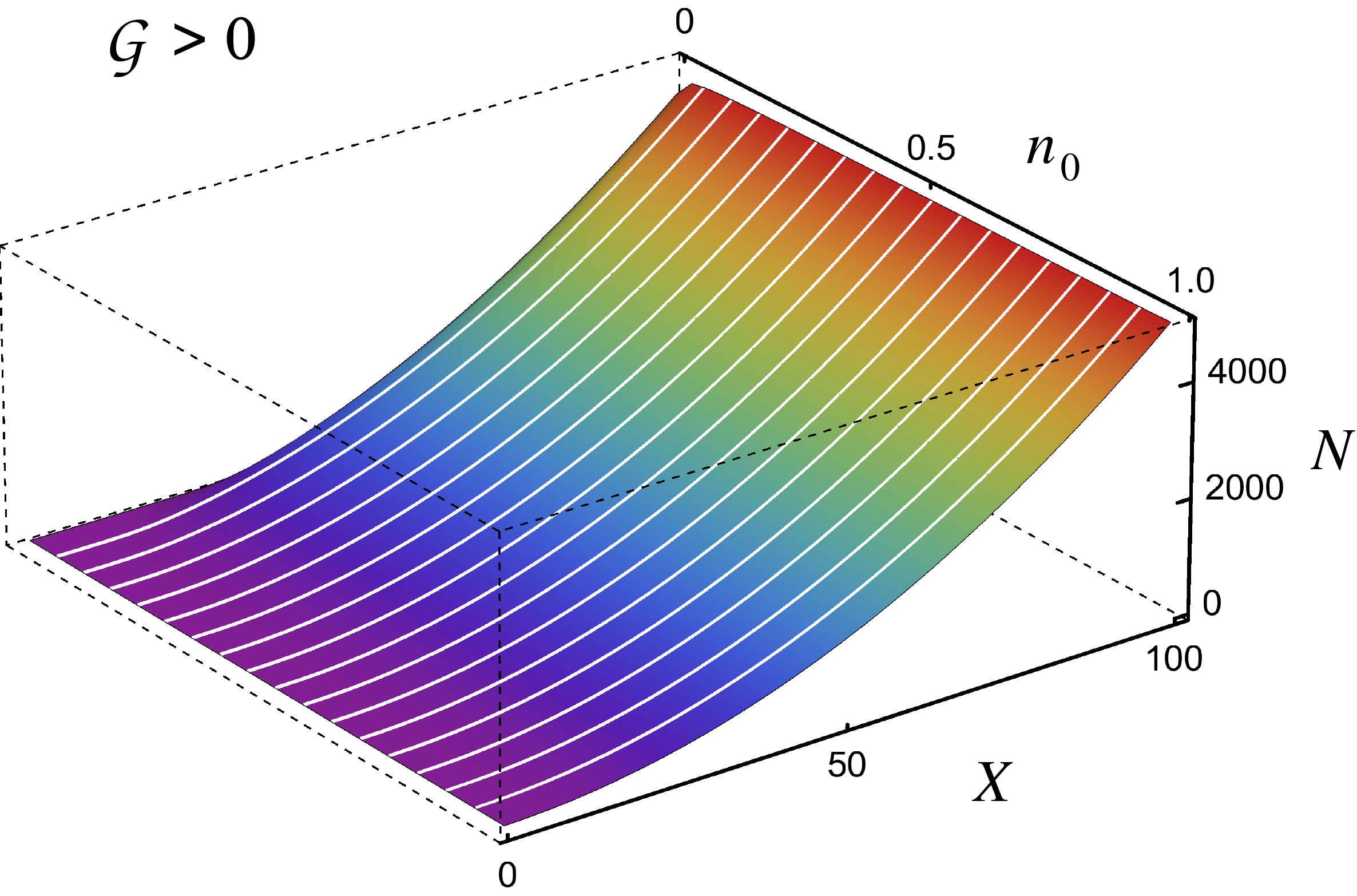} 
  \caption{The function $N(X;n_0)$, i.e., the total amount of
    ``hard--core spheres'' inside a circular region of given radius, for
    ``attractive''  ($\qmar < 0 $, left panel) or ``repulsive'' 
    ($\qmar > 0 $, right panel) Marangoni flow. 
   Each white line corresponds to a curve $N(X; n_0)$ for
      constant $n_0$. 
    }
  \label{fig:hardN}
\end{figure*}

The function $N(X; n_0)$ given by \eq{eq:N} has been evaluated for
each density profile, identified by the central density $n_0$. The
results are shown in \fig{fig:hardN}. The line $N(X;1)=X^2/2$
(best visible in the case of $\qmar < 0$ as the border of the white
region in the back vertical plane) represents the border of the
physically accessible region, i.e., beyond which the total packing
fraction is higher than close packing (one cannot pour in more hard
particles than actually fit in the region). The curves $N(X; n_0)$
seem to cover completely the interior of the physically accessible
region without mutual crossing. Thus, the relationship between the
central density $n_0$ and the pair $(X,N)$ is one-to-one in the
explored range of values.

\subsection{Experimental relevance}

We now consider the theoretical predictions derived in the previous
Sections from the perspective of potential experimental validation.
For that purpose, we focus on the more interesting case of the soft repulsive 
interaction and provide numerical estimates for the various
parameters appearing in the theoretical model. Additionally, this provides the 
means to critically assess the validity of the assumptions involved in the
model.

In order to translate the results derived in the previous Sections
into physical units, we have to compute the characteristic density
scale $\varrho_c=\zeta^{-2}$ and the characteristic length scale
$\ell=1/\sqrt{|g| \varrho_c}=\zeta/\sqrt{|g|}$, see
\eq{eq:lengthscale}. More precisely, since \eqs{eq:model} provide a
coarse--grained description, one would ideally require that the
characteristic length $\ell$ is much larger than the mean
interparticle separation, i.e.,
  \begin{equation}
    \label{eq:largeell}
    \ell \sqrt{\varrho} = \frac{\ell \sqrt{n}}{\zeta} = \sqrt{\frac{n}{|g|}} \gg 1 .
  \end{equation}
  Additionally, 
    the model equations have been obtained in the
  dilute approximation; this requires 
  that the mean interparticle separation is much larger than the
  radius $R$ of the particles, i.e.,
  \begin{equation}
    \label{eq:largezeta}
    \frac{1}{R\sqrt{\varrho}} = \frac{\zeta}{R\sqrt{n}} \gg 1 .
  \end{equation}
   These two constraints set the limits of applicability
   of the theoretical predictions. Although they depend locally on the
   value of the density $n$, they do so weakly; thus, for the purpose
   of getting order-of-magnitude estimates, one can simply set $n=1$
   in \eqs{eq:largeell} and (\ref{eq:largezeta}).
   Therefore, \eq{eq:largeell} states that the activity cannot be too
   large, while \eq{eq:largezeta} states that the repulsion between
   the soft spheres cannot be too weak.

The value of the parameter $\zeta$, which sets the mean
  interparticle separation, depends on the specific physical origin of
  the repulsion given by \eq{eq:Vdip}. We consider here its value in
  some representative physical systems that have been previously studied:
  \begin{enumerate}
  \item In the experimental setup with paramagnetic particles reported
    in \rcite{ZLM99}, it is (see \rcite{DOD10})
    \begin{equation}
      \label{eq:zetaH}
      \frac{\zeta}{R} \approx 2.30 \left(\frac{H}{ 
          \mathrm{hA/m}}\right)^{2/3} \frac{R}{\mathrm{\mu m}} ,
    \end{equation}
    with an applied magnetic field in the range $H\approx
    1$--$10\; \mathrm{hA/m}$ ($\mathrm{hA}=10^2\;\mathrm{A}$ is
    hecto-ampere; the lowest bound is set by Earth's magnetic field of
    strength $\approx 0.4\;\mathrm{hA/m}$).
  \item For the case of particles polarized by an external electric
    field $E$, one 
    has (see \rcite{DOD10})
    \begin{equation}
      \label{eq:zetaE}
      \frac{\zeta}{R} \approx 3.74 \left(\frac{E}{ 
          \mathrm{MV/m}}\right)^{2/3} \frac{R}{\mathrm{\mu m}} ,
    \end{equation}
    when the parameter values of the experiment reported in
    \rcite{ASJN08} are used (here $\mathrm{MV}=10^6\;\mathrm{V}$ is
    mega-volt).
  \item Polymeric or dielectric oxide (e.g., silica, titania, alumina)
    colloidal particles immersed in water generically become ionized.
    When, additionally, they are trapped at the interface between
    water and a dielectric fluid (e.g., air or oil), their mutual
    interaction takes the form of \eq{eq:Vdip}. For polystyrene
    particles at a water--oil interface, as in the experimental
    configuration employed in \rcite{PLMB15}, and assuming weakly
    charged colloids, we derive the estimate (see
    Appendix~\ref{app:zetaion})
    \begin{equation}
      \label{eq:zetaion}
      \frac{\zeta}{R} \approx 0.59 
      \left(\frac{\mathrm{M}}{c} \, \frac{R}{\mathrm{\mu m}}\right)^{1/3},
    \end{equation}
    where $c$ is the molar (M) concentration of the monovalent ions in water. 
    If the charge of the
    particles is very large, nonlinear screening effects dominate and
    the estimate~(\ref{eq:zetaion}) is invalid because the Bjerrum
    length becomes almost independent of the ionic concentration of the solution
    \cite{FDO07}.  
\end{enumerate}

The parameter $\qadim$ gives the magnitude of the Marangoni flow, 
Eqs.~(\ref{eq:u}) and (\ref{eq:smallq}).
Its sign will depend on how the surface tension of the interface is influenced 
by the chemical or thermal activity of the particles and on whether the particles are 
sources or sinks of that tensioactive component.
Thus, one gets an ``attractive'' Marangoni flow ($\qadim<0$) for
chemically active particles either (i) if the surface tension
decreases with increasing concentration of the chemicals and the
latter are released by the particles or (ii) if the surface tension
increases and the chemicals are adsorbed by the particles
\cite{MaSh14,DMPD16a}. For thermally active particles, this
``attractive'' effect is achieved when either (i) the particles are
sources of heat and the surface tension is reduced with increasing
temperature, or (ii) the particles absorb heat and the surface tension
grows with temperature \cite{LGN97,Wuer14}. The trends must be
reversed in order to achieve a ``repulsive'' Marangoni flow
($\qadim>0$).

In order to be definite, one can estimate $g$ for the cases of
activity that have been discussed in experimental reports. Thus, for
platinum--covered active particles catalysing the decomposition of
water peroxide we get (see App.~\ref{app:marang})
\begin{equation}
  \label{eq:qPt}
  |\qadim| \approx 0.075 \left(\frac{R}{\mathrm{\mu m}}\right)^3 ,
\end{equation}
when an air--water interface is considered. For an interface between
two dense fluids (like water and oil, as in the experiments reported
in \rcite{Isa2017}), this value is expected to be enhanced by up to
four orders of magnitude due to the reduced diffusivity of molecular
oxygen in the liquid phases compared to the one in a gas phase
\cite{DMPD16a}. For particles that become thermally active due to
heating by a laser, the value of $g$ for micron--sized particles is
larger by over six orders of magnitude (see App.~\ref{app:marang}).

\begin{figure*}
  \centering 
  \includegraphics[width=.45\textwidth]{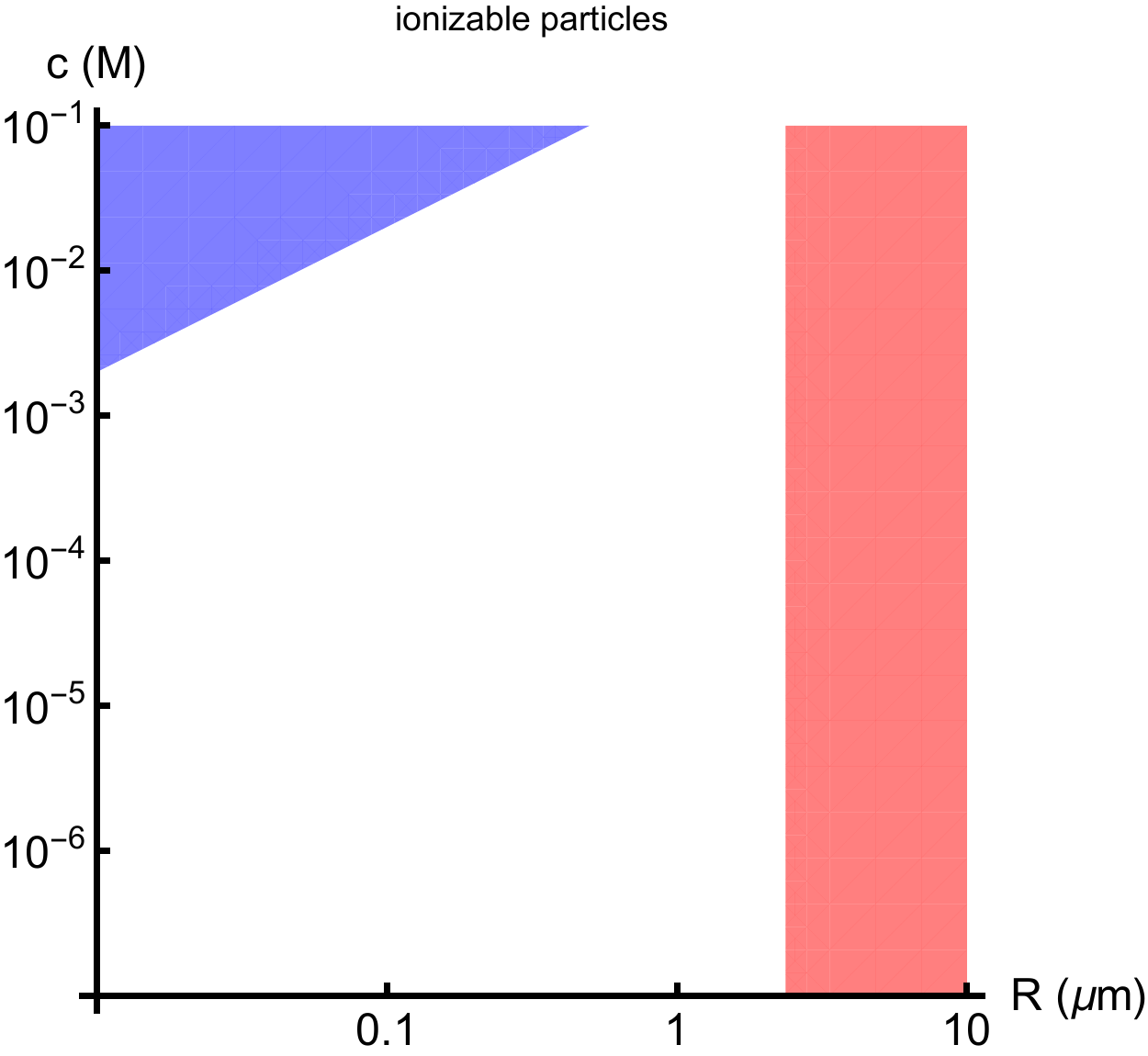}
  \hfill
  \includegraphics[width=.45\textwidth]{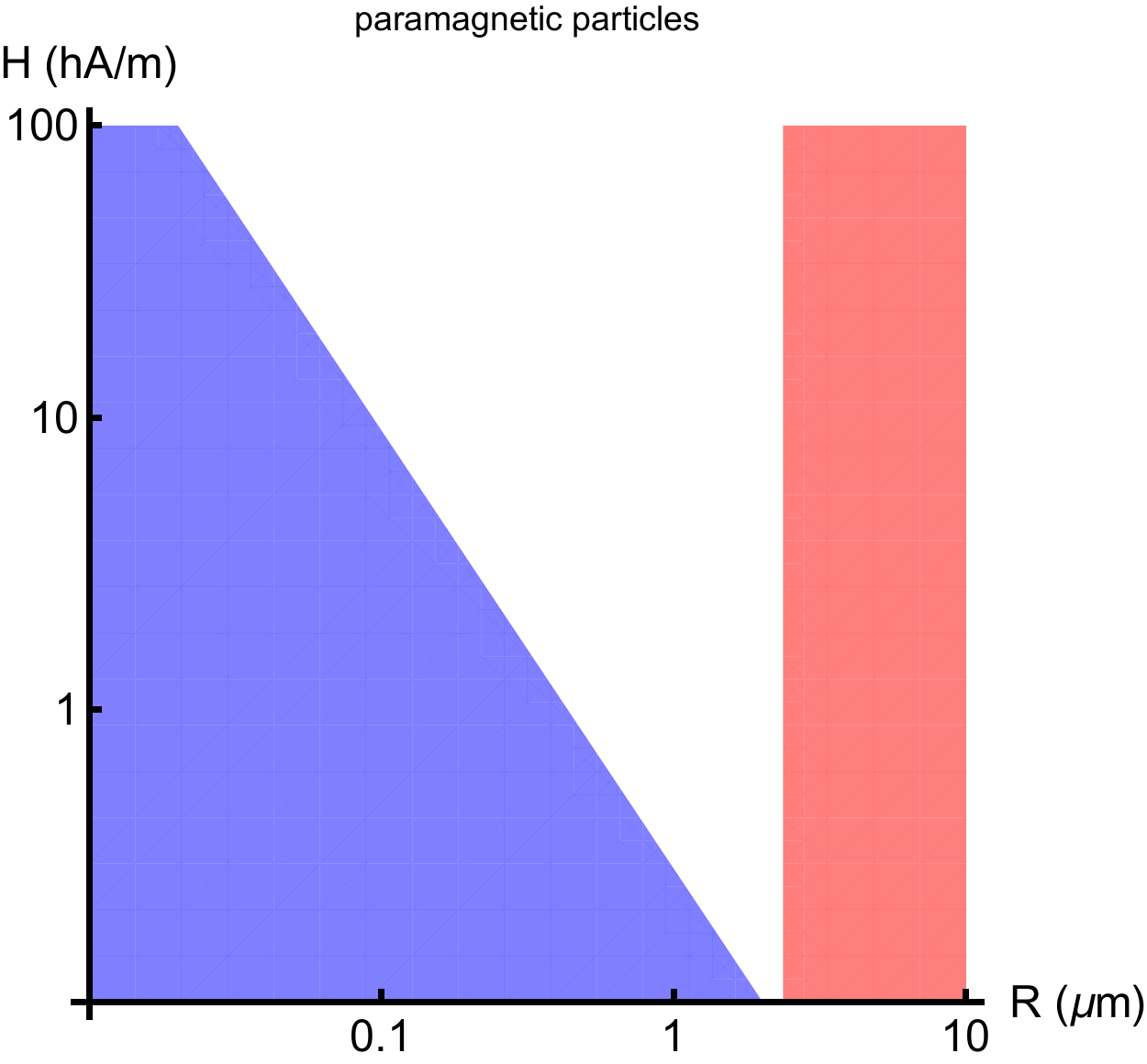}
  \caption{The parameter space for different experimental
    realizations. The unshadowed area is the parameter region where
    the constraints given by Eqs.~(\ref{eq:largeell}) and
    (\ref{eq:largezeta}) (with $n=1$) are expected to hold. The
      red area represents the region where $\ell/\zeta<1$, the
      activity $g$ being estimated with \eq{eq:qPt}. In the blue area
      it is $\zeta/R<1$, 
      with $\zeta$ estimated from \eq{eq:zetaion} for the left
      plot, and from \eq{eq:zetaH} for the right one.
    }
  \label{fig:constraints}
\end{figure*}

By combining this estimate of $\qadim$ with any of the previous
estimates of the Bjerrum length $\zeta$, one can
check the fulfillment of the constraints~(\ref{eq:largeell}) and
(\ref{eq:largezeta}). This is illustrated in \fig{fig:constraints},
which shows the region in parameter space where the constraints are
expected to hold. The left plot corresponds to ionizable particles,
with $\zeta$ given by \eq{eq:zetaion}; the right plot to paramagnetic
particles with $\zeta$ given by \eq{eq:zetaH}. (The plot obtained with
the estimate~(\ref{eq:zetaE}) for electrically polarized particles is
essentially indistinguishable from the right plot when the vertical
axis is reinterpreted as electric field measured in $\mathrm{MV/m}$.)

We remark here that we are not in a position to tackle the specific
details of an actual experimental realization, for instance, how a
platinum--covered active particle can be fabricated that is
paramagnetic or remains ionizable. Thus, the results shown in
Fig. \ref{fig:constraints} have to be understood as a rough guide for
the search of an optimal experimental configuration. In this regard,
the plots suggest that there are experimentally accessible ranges of
parameter values where the model is valid and the theoretical
predictions can be tested.  Consider, for instance, a collection of
ionizable particles of radius $R=0.5\;\mathrm{\mu m}$ at a water--oil
interface with $c=10^{-6}\;\mathrm{M}$ (for pure water). In this case,
$\zeta\approx 46 R$, and thus the dilute approximation is well
grounded. The corresponding characteristic length $\ell$ is also
large, i.e., $\ell \approx 10\zeta$, and thus the full ``onion-like''
structure of crystalline core, fluid phase, and hexatic transition
region, see Figs.~\ref{fig:softprofile} and \ref{fig:Xfreeze}, should
be experimentally accessible. Similarly, paramagnetic particles of the
same radius at a water--air interface in a large magnetic field
$H=10^2\;\mathrm{hA/m}$ would give the values $\zeta\approx 25 R$ and
$\ell \approx 10\zeta$; such system thus provides another potential
candidate for an experimental validation of the theoretical
predictions.

We finally notice that the flow given by \eq{eq:radialu} would likely
be not directly observable. The characteristic velocity scale is
$\Gamma kT/\ell$, as derived from \eqs{eq:cumulativemass}. For the
examples considered above, this scale lies in the range of nanometers
per second. Therefore, the effect of the activity of the particles
would not be so much apparent in the induced Marangoni flow as in the
``onion-like'' structure of phase coexistences.

\section{Conclusions}
\label{sec:conclusions}

We have studied a mean--field model for the phase coexistence in a
monolayer of colloidal particles with repulsive interactions, located
at a fluid interface, and experiencing Marangoni flow induced by the
activity of the particles. We have analyzed in detail the case of
``soft'' (decaying as $r^{-3}$) and hard--core repulsive interactions.
We have found stationary states with an ``onion-like'' structure in
the particle distribution within the monolayer.  For the most relevant
case of soft repulsion, the density reaches the values corresponding
to the liquid to hexatic (freezing) and solid to hexatic (melting)
transitions in two dimensions, respectively. Furthermore, we have
thoroughly and critically discussed the relevance of these results for
potential experimental realizations and concluded that experimental
validation of the theoretical predictions seems feasible.

A further interesting consequence of the constraints expressed by
Eqs. (\ref{eq:largeell}) and (\ref{eq:largezeta}) is that they lead to
opposing restrictions on the size of the colloidal particles, with the
radius $R$ being typically bracketed between a maximum and a minimum
value (see Fig.~\ref{fig:constraints}). A weak Marangoni flow expands
the range of values of $R$ for validity of the theoretical results. It
was noticed in \rcite{DMPD16a} that the small value
$g\sim 0.01$--$0.1$ in usual configurations could explain the lack of
experimental observations of the activity--induced Marangoni
flow. However, the Marangoni flow has a long--ranged component
formally analogous to gravity. As a consequence, these lead to the
interesting aspect that, although the drag velocity by the Marangoni
flow may be unobservably small, the added-up effect of many particles
yields noticeable collective effects (see, e.g., the qualitative discussion on this
issue in \rcite{DOD10}). Therefore, the most relevant conclusion of
this work is that the spatially structured phase coexistence \emph{is}
the observationally accessible signature of the activity--induced
Marangoni flow.

When this flow is too strong, the constraint in \eq{eq:largeell} is
not satisfied, which usually hints that the mean--field
expression~(\ref{eq:u}) is a poor approximation and that interparticle
correlations have to be accounted for in the computation of the
Marangoni flow. Likewise, violation of the constraint in
\eq{eq:largezeta} indicates that hydrodynamical interactions other
than the Marangoni flow, here including short--ranged corrections,
have to be considered. Both scenarios call for a cautionary use of
the notion of ``phase coexistence'', since a phase is a well--defined
concept in equilibrium thermodynamics only. Its use in the context of
the addressed problem is meaningful thanks to the wide separation of
length scales expressed by Eqs.~(\ref{eq:largeell}) and
(\ref{eq:largezeta}): in such case, one can apply the hypothesis of
local equilibrium in order to comprise the relevant contribution by
the interparticle direct forces into the equilibrium chemical
potential $\mu(\varrho)$. This latter function then provides the
ground for the notion of a local ``thermodynamic phase''.

When the length scales are not well separated, it may not be
appropriate to interpret the features of a stationary density profile
in terms of thermodynamic phases. This, however, opens up the
possibility of an exciting new scenario, namely that those features
actually represent the nonequilibrium counterpart of ``phases'' which
are induced by the hydrodynamic interactions. For instance, the
crystal--gas profiles derived for hard--spheres (see
Fig.~\ref{fig:hardprofile}) clearly violate the dilute--limit
approximation and the theoretical model must be modified in order to
provide reliable predictions. Thus, the analysis presented in
Sec.~\ref{sec:hard} should be viewed as a first step towards a more
realistic model that accounts for the changes in the
``crystal--gas--coexistence'' interpretation which would be induced by
the short--ranged hydrodynamic interactions.

\appendix

\section{Appendix: equation of state for soft spheres}
\label{app:musoft}

Under isothermal conditions, the chemical potential of the particles 
in the monolayer, $\mu(\varrho,T)$, can be related to the equation of state 
$p(\varrho,T)$ for the pressure by means of the thermodynamic identity 
\begin{equation}
\label{eq:mu_der}
  \left(\frac{\partial\mu}{\partial \varrho} \right)_T 
  = \frac{1}{\varrho} \left( \frac{\partial p}{\partial
      \varrho}\right)_T .
\end{equation}
The evaluation of the partition function for a collection a
  particles interacting with the potential given by \eq{eq:Vdip}
  straightforwardly leads to the scaling relationship   $p(\varrho,T) = k T \zeta^{-2} 
\hat{p}(n)$ with $n=\varrho
  \zeta^2$. Then, the expression~(\ref{eq:mu_der}) renders
  $\mu(\varrho,T)=k T \hat{\mu}(n)$. This scaling (which is shared with the
  hard spheres case) shows that the collection of soft spheres is an
  athermal system, i.e., its phase diagram depends only on density,
  not on temperature.

The equation of state $\hat{p}(n)$ can be determined by a
combination of numerical results and theoretical estimates. Thus,
Monte Carlo simulations \cite{DOD10} show that the following fit holds
to a good approximation in the high density range of the fluid phase:
\begin{equation}
 \hat{p}_\mathrm{fluid} (n) = n (3 + 6.6 n^{3/2}) ,
  \qquad
  2 \lesssim n < n_\mathrm{freeze},
\end{equation}
where $n_\mathrm{freeze} \approx 4.65$ is the density at the freezing transition. In the 
solid phase, theoretical arguments confirmed by comparison with numerical results 
\cite{Spee03} suggest a fitting of the form
\begin{equation}
   \hat{p}_\mathrm{solid} (n,T) = n \left( \frac{5}{2} + K n^{3/2} \right) ,
  \qquad
  n_\mathrm{freeze} < n .
\end{equation}
with a certain constant $K$. This analysis neglects the existence of a
hexatic phase between the liquid and the solid ones, given that the
range of existence is very narrow
($n_\mathrm{melt} - n_\mathrm{freeze} \approx 0.22$). One can provide an
estimate of the value of $K$ by imposing \textit{ad hoc} the condition
that the fluid and solid phases coexist at the freezing density:
\begin{equation}
  \hat{p}_\mathrm{fluid}(n_\mathrm{freeze}) = \hat{p}_\mathrm{solid}(n_\mathrm{freeze}) 
  \qquad\Rightarrow\qquad
  K \approx 6.65 .
\end{equation}
Since $\hat{p}_\mathrm{solid} > 442$, the relative error committed in
estimating the pressure $\hat{p}_\mathrm{solid}$ of the solid phase
with the fitting formula $\hat{p}_\mathrm{fluid}$ for the
high--density fluid is less than $1\%$. Therefore, it can be assumed
\begin{equation}
  \label{eq:approxp}
  \hat{p} (n) = n (3 + 6.6 n^{3/2}) ,
  \qquad
  2 \lesssim n ,
\end{equation}
in the whole range of high densities, regardless of the phase being
considered. For low densities, $n<2$, this fit departs from the
  values derived from simulations. However, the deviations are not very large (e.g., 
  less than $25\%$ if $n>0.5$) and, indeed, we have
  checked that the profiles shown in \fig{fig:softprofile} are
  practically indistinguishable when either the simulation data or the
  extrapolation of \eq{eq:approxp} down to $n\to 0$ are used to compute
  them. Therefore, for our purposes one can use the chemical potential
obtained by integrating \eq{eq:mu_der} with \eq{eq:approxp} for the
whole range of densities considered in this work:
\begin{equation}
  \label{eq:fitmusoft}
  \hat{\mu}(n) = 3 \ln n + 11 n^{3/2} + \mu_0 ,
\end{equation}
with an irrelevant integration constant $\mu_0$.

\section{Appendix: equation of state for hard spheres}
\label{app:muhard}

The equation of state of hard spheres exhibits the simple scaling
$p(\varrho,T)=k T \varrho_c \hat{p}(n)$, with $n=\varrho/\varrho_c$ in
terms of the close packing density $\varrho_c$. This carries over to a
scaling $\mu(\varrho,T)=k T \hat{\mu}(n)$ for the chemical potential
via \eq{eq:mu_der}. Thus, the monolayer of hard spheres is an athermal
system and its phase behavior depends only on density. For our
purposes, it is sufficient to use the approximate equation of
state of a two--dimensional fluid of hard spheres (i.e., hard disks)
provided by \rcite{GZB97}:
\begin{equation}
  \hat{p} (n) = n \frac{1+n}{1-n} ,
\end{equation}
with $\varrho_c=1/(2\sqrt{3} R^2)$ denoting the close packing density
for disks of radius $R$. This leads to the chemical potential
\begin{equation}
  \label{eq:fitmuhard}
  \hat{\mu}(n) = \ln\frac{n}{(1-n)^2} + \frac{2}{1-n} + \mu_0,
\end{equation}
with an irrelevant integration constant $\mu_0$.

\section{Appendix: Bjerrum length for ionizable particles}
\label{app:zetaion}

When ionizable particles are trapped at the interface between
  water and a dielectric fluid, they experience a mutual long--ranged
  repulsion because, unlike in bulk water, the electric field is not
  completely screened. This repulsion follows \eq{eq:Vdip} because
  each particle can be modelled as an electric dipole perpendicular to
  the interface \cite{Stil61,Hurd85,DFO08}. Although this expression is
  justified in principle only as an asymptotic approximation between
  sufficiently distant pairs, its validity has been confirmed
  experimentally \cite{ACNP00,PLMB15}. Of particular interest for our
  purposes is that the interaction can be actually described well with
  \eq{eq:Vdip} for a wide range of monolayer densities that includes
  the ones of the transitions involving the hexatic phase \cite{PLMB15,KGHC17}.
  
  If the total charge $q$ of the particle is sufficiently small,
  linear screening holds and the strength of the dipole is
  proportional to the product of the charge $q$ with the Debye length
  $\kappa^{-1}$ in water \cite{Stil61,Hurd85}. As a consequence, the
  proportionality constant $B$ in \eq{eq:Vdip} satisfies the scaling
  $B\propto (q/\kappa)^2$. For a symmetric electrolyte, the Debye
  length depends on the concentration $c$ of monovalent ions as
  $\kappa\propto\sqrt{c}$. Assuming that the charged acquire by the
  particle is proportional to its surface, one arrives at the relation
  \begin{equation}
    \label{eq:Bion}
    B = \hat{B} \frac{R^4}{c} .
  \end{equation}
  The constant $\hat{B}$ can be estimated, for instance, from the
  experimental study described in \rcite{PLMB15}: the value
  $B/kT \approx 2\times10^5 \,\mathrm{\mu m}^3$ is measured, which is
  of the same order of magnitude as previously reported values, when
  using polystyrene particles of radius $R=1\;\mathrm{\mu m}$ trapped
  at an interface between oil and very pure water, for which we take
  the conservative estimate $c=10^{-6}\;\mathrm{M}$. The expression
  in \eq{eq:Bion} then leads to the estimate, \eq{eq:zetaion}, for the
  Bjerrum length.

\section{Appendix: strength of the Marangoni flow}
\label{app:marang}

In order to get a quantitative estimate of the strength $g$ of the
Marangoni flow, we start from the detailed calculation in the
point--particle (monopolar) approximation. For a chemically active
particle, the computation shows that the magnitude of the Marangoni
flow, as appears in \eq{eq:u}, is \cite{DMPD16a}
\begin{equation}
\label{eq:Qdef}
 \qmar = \frac{Q b_0}{8 D_{+} \eta_{+}} .
\end{equation}
Here, $Q$ denotes the source/sink (monopole) strength of the active
particle, i.e., the amount of chemical released/absorbed by an active
a particle per unit time. $b_0$ denotes the coefficient of the linear
response of the surface tension to changes in the concentration of
chemical species or in temperature at the interface, while $\eta_{+}$
and $D_{+}$ are average values of the viscosity and of the diffusion
constant of chemicals in the two fluids, respectively.  The expression
for the case of a thermally active particle is exactly the same
\cite{LGN97,Wuer14} with the reinterpretation of $Q$ as the heat
exchanged per unit time between the particle and the fluids, of $b_0$
as the surface tension response to temperature changes, and of $D_+$
in terms of the heat conductivities of the fluids.

In the dilute limit, which is implicit in our model, the mobility
$\Gamma$ can be approximated by the Stokes formula for a sphere of
radius $R$ in a fluid of viscosity $\eta_{+}$. Then, from
Eqs.~(\ref{eq:smallq}) and (\ref{eq:Qdef} one arrives at
\begin{equation}
\label{eq:q_estimate}
\qadim = \frac{3 \pi Q b_0 R}{4 D_+ k T} .
\end{equation}
The dependences on the details of the activity and on the effect of
the active component on the surface tension makes difficult to
estimate the value of $\qadim$. We have thus considered for
definiteness two specific configurations, for chemically or thermally
active particles, respectively:
\begin{enumerate}
\item For chemically active particles, we note that $\qadim$ in
  \eq{eq:q_estimate} is a factor of 25 larger than the similar
  quantity $q$ estimated in \rcite{DMPD16a}, so that one can
  straightforwardly adapt the available estimates for $q$ to the
  problem of interest here. Thus, for an active particle made of
  platinum and decomposing hydrogen peroxide dissolved in the fluids,
  for which $Q/(4\pi R^2) \approx 10^{-3}\;\mathrm{mol/(s\times m^2)}$
  in the experiment described in \rcite{Paxton2004}, one obtains the
  estimate quoted in \eq{eq:qPt} for an air--water interface,
  corresponding to the values $D_+\sim 10^{-4} \;\mathrm{m^2/s}$,
  $\eta_+\sim 10^{-3}\;\mathrm{Pa\times s}$,
  $b_0\sim 10^{-3}\;\mathrm{N/(m\times M)}$.

\item For thermally active particles, we refer to the simulations
  reported in \rcite{BVKV12} of a self--propelled particle due to it
  being heated by a laser. For micron--sized particles, this provides
  a temperature gradient of the order of $1\;\mathrm{K/\mu m}$ in the
  fluid close to the particle. Since the temperature contrast
  $\Delta T(r)$ at a distance $r$ from the particle is given as
  $\Delta T = Q/(4\pi D_+ r)$ in the monopolar approximation
  \cite{Wuer14,DMPD16a}, we estimate
  $Q/(4\pi D_+) \approx 1\;\mathrm{K\times\mu m}$. Therefore, for
  $R=1\;\mathrm{\mu m}$ and
  $b_0\approx -2\times 10^{-4}\;\mathrm{N/(m\times K)}$ for an
  air--water interface, \eq{eq:q_estimate} gives a value of $g$ which
  is $10^6$ times larger than the case of the chemically active
  particle.
\end{enumerate}



\section{Acknowledgements}

A.D. acknowledges support by the Ministerio de Econom{\'i}a y
Competitividad del Gobierno de Espa{\~n}a through Grant
FIS2017-87117-P (partially financed by the European Regional
Development Fund). This article is based upon work from COST Action
MP1305 ``Flowing Matter'', supported by COST (European Cooperation in
Science and Technology). The authors thank Martin Oettel for providing 
the MC simulations data.



\balance


\bibliographystyle{rsc} 

\providecommand*{\mcitethebibliography}{\thebibliography}
\csname @ifundefined\endcsname{endmcitethebibliography}
{\let\endmcitethebibliography\endthebibliography}{}

\end{document}